# Conception of the scalar-vector potential in contemporary electrodynamics

F.F. Mende
E – mail: mende@mende.ilt.kharkov.ua

B.I. Verkin Institute for Low Temperature Physics and Engineering, NAS

Ukraine, 47 Lenin Ave., Kharkov, 61164, Ukraiua

## Abstract

The present paper is devoted to consideration and discussion of a number of contradictions that take place in fundamental electrodynamics researches. A concept of the scalar-vector potential is introduced that allows us to avoid a number of challenges taking place in treatment of forces of elecrtromagnetic interaction. It is introduced the delayed scalar-vector potential allowing one to solve radiation problems at an elementary level. It is shown that in a nonmagnetized plasma, beside the earlier known longitudinal Langmuir resonance, there may exist a resonance transversal relative to the wave propagation direction. It is also shown that the dielectric and magnetic permittivities are independent on the frequency. It is introduced a concept of the kinetic capacitance.

## Introduction

We got used to such a concept as magnetic field insomuch that could not imagine contemporary electrodynamics without it. This habit forced us to blink at the obvious imperfections of this theory. The concept of magnetic field assumes the presence of magnetic forces only between charges that move relative to an immobile reference system. It is supposed that some moving charges induce a magnetic field and some ones interact with this field. From this it follows that two charges moving in parallel must attract one another. However, if one turns to another reference system moving together with the charges then there are no magnetic fields in such a system. It should be noted that the contradiction considered can not be eliminated neither in the frame of classical electrodynamics, nor in the frame of special theory of relativity.
    Another disconformity consists in that the forces acting between current-carrying systems, as well as poderomotive forces are applied not to moving charges but to the ion lattice. In the concept of magnetic field there are no directions why the force impact is exerted on the lattice but not the moving charges.

Unfortunately, nobody pays attention some how to these obvious discomformities. At the same time, neglecting these facts recoils negatively upon solving a number of physical and technical problems. The main hopes of implementation of controlled thermonuclear synthesis are currently associated with employment of the pinch-effect, however, we have no clear notion of why not only moving electrons undergo the contraction but also ions.

# 1. Equations of electromagnetic induction in moving coordinates

The Maxwell equations do not permit us to write down the fields in moving coordinates proceeding from the known fields measured in the stationary coordinates. Generally, this can be done through the Lorentz transformations but they so not follow from classical electrodynamics. This brings up the question: Can classical electrodynamics furnish correct results for the fields in a moving coordinate system, or at least offer an acceptable approximation? If so, what form will the equations of electromagnetic induction have?

The Lorentz force is

$$\vec{F}' = e\,\vec{E} + e\,[\vec{V} \times \vec{B}]. \qquad (1.1)$$

It bears the name of Lorentz it follows from his transformations which permit writing the fields in the moving coordinates if the fields in the stationary coordinates are known. Henceforward, the fields and forces generated in a moving coordinate system will be indicated with primed symbols.

The clues of how to write the fields in moving coordinates if they are known in the stationary system are available even in the Faraday law. Let us specify the form of the Faraday law:

$$\oint \vec{E}'\,d\,\vec{l}' = -\frac{d\,\Phi_B}{d\,t}. \qquad (1.2)$$

The specified law, or, more precisely, its specified form, means that $\vec{E}$ and $d\vec{l}$ should be primed if the contour integral is sought for in moving coordinates and unprimed for stationary coordinates. In the latter case the right-hand side of Eq. (1.2) should contain a partial derivative with respect to time which fact is generally not mentioned in literature.

The total derivative with respect to time in Eq. (1.2) implies that the final result for the contour e.m.f. is independent of the variation mode of the flux. In other words, the flux can change either purely with time variations of $\vec{B}$ or because the system, in which $\oint \vec{E}\,d\vec{l}'$ is measured, is moving in the spatially varying field $\vec{B}$. In Eq. (1.2)

$$\Phi_B = \int \vec{B}\,d\,\vec{S}', \qquad (1.3)$$

where the magnetic induction $\vec{B} = \mu \vec{H}$ is measured in the stationary coordinates and the element $d\vec{S}'$ in the moving coordinates.

Taking into account Eq. (1.3), we can find from Eq. (1.2)

$$\oint \vec{E}' d\vec{l}' = -\frac{d}{dt} \int \vec{B} \, d\vec{S}'. \tag{1.4}$$

Since $\dfrac{d}{dt} = \dfrac{\partial}{\partial t} + \vec{V} \, grad$, we can write

$$\oint \vec{E}' d\vec{l}' = -\int \frac{\partial \vec{B}}{\partial t} d\vec{S} - \int [\vec{B} \times \vec{V}] d\vec{l}' - \int \vec{V} \, div \, \vec{B} \, d\vec{S}'. \tag{1.5}$$

In this case contour integral is taken over the contour $d\vec{l}'$, covering the space $d\vec{S}'$. Henceforward, we assume the validity of the Galilean transformations, i.e. $d\vec{l}' = d\vec{l}$ and $d\vec{S}' = d\vec{S}$. Eq. (1.5) furnishes the well-known result:

$$\vec{E}' = \vec{E} + [\vec{V} \times \vec{B}], \tag{1.6}$$

which suggests that the motion in the magnetic field excites an additional electric field described by the final term in Eq. (1.6). Note that Eq. (1.6) is obtained from the slightly specified Faraday law and not from the Lorentz transformations.

According to Eq. (1.6), a charge moving in the magnetic field is influenced by a force perpendicular to the direction of the motion. However, the physical nature of this force has never been considered. This brings confusion into the explanation of the homopolar generator operation and does not permit us to explain the electric fields outside an infinitely long solenoid on the basis of the Maxwell equations.

To clear up the physical origin of the final term in Eq. (1.6), let us write $\vec{B}$ and $\vec{E}$ in terms of the magnetic vector potential $\vec{A}_B$:

$$\vec{B} = rot \, \vec{A}_B, \qquad \vec{E} = -\frac{\partial \vec{A}_B}{\partial t}. \tag{1.7}$$

Then, Eq. (1.6) can be re-written as

$$\vec{E}' = -\frac{\partial \vec{A}_B}{\partial t} + [\vec{V} \times rot \, \vec{A}_B], \tag{1.8}$$

and further:

$$\vec{E}' = -\frac{\partial \vec{A}_B}{\partial t} - (\vec{V} \, \nabla)\vec{A}_B + grad\left(\vec{V} \, \vec{A}_B\right). \tag{1.9}$$

The first two terms in the right-hand side of Eq. (1.9) can be considered as the total derivative of the vector potential with respect to time:

$$\vec{E}' = -\frac{d\vec{A}_B}{dt} + grad\left(\vec{V}\vec{A}_B\right). \qquad (1.10)$$

As seen in Eq. (1.9), the field strength, and hence the force acting upon a charge consists of three components.

The first component describes the pure time variations of the magnetic vector potential. The second term in the right-hand side of Eq. (1.9) is evidently connected with the changes in the vector potential caused by the motion of a charge in the spatially varying field of this potential. The origin of the last term in the right-hand side of Eq. (1.9) is quite different. It is connected with the potential forces because the potential energy of a charge moving in the potential field $\vec{A}_B$ at the velocity $\vec{V}$ is equal to $e\left(\vec{V}\vec{A}_\mathbf{B}\right)$. The magnitude $e\,grad\left(\vec{V}\vec{A}_B\right)$ describes the force just as the scalar potential gradient does.

Using Eq. (1.9), we can explain physically all the strength components of the electronic field excited in the moving and stationary cooperates. If our concern is with the electric fields outside a long solenoid, where the no magnetic field, the first term in the right-hand side of Eq. (1.9) come into play. In the case of a homopolar generator, the force acting upon a charge is determined by the last two terms in the right-hand side of Eq.(1.9), both of them contributing equally.

It is therefore incorrect to look upon the homopolar generator as the exception to the flow rule [1-2] because, as we saw above, this rule allows for all the three components. Using the rotor in both sides of Eq. (1.10) and taking into account $rot\,grad \equiv 0$, we obtain

$$rot\,\vec{E}' = -\frac{d\vec{B}}{dt}. \qquad (1.11)$$

If motion is absent, Eq. (1.11) turns into Maxwell equation (1.2). Equation (1.11) is certainly less informative than Eq. (1.2): because of $rot\,grad \equiv 0$, it does not include the forces defined in terms of $e\,grad\left(\vec{V}\vec{A}_B\right)$. It is therefore more reasonable to use Eq. (1.2) if we want to allow for all components of the electric fields acting upon a charge both in the stationary and in the moving coordinates.

As a preliminary conclusion, we may state that the Faraday Law, Eq. (1.2), when examined closely, explains clearly all features of the homopolar generator operation, and this operation principle is a consequence, rather than an exception, of the flow rule, Eq. (1.2). Feynman's statement [2] that $\left[\vec{V}\times\vec{B}\right]$ for the "moving contour" and $\nabla\times\vec{E} = -\frac{\partial\vec{B}}{\partial t}$ for the "varying field" are absolutely different laws is contrary to fact. The Faraday law is just the sole unified fundamental principle which Feynman declared to be missing. Let us clear up another Feynman's interpretation. Faraday's observation in fact led him to discovery of a new law relating electric and magnetic fields in the region where the magnetic field varies with time and thus generates the electric field. This correlation is essentially true but not complete. As shown above, the electric field can also be excited where there is no magnetic field, namely, outside an infinitely long solenoid. A more complete formulation follows from Eq. (1.9) and the relationship $\vec{E} = -\dfrac{d\vec{A}_B}{dt}$ is more general than $rot\,\vec{E} = -\dfrac{\partial \vec{B}}{\partial t}$.

This suggests that a moving or stationary charge interacts with the field of the magnetic vector potential rather than with the magnetic field. The knowledge of this potential and its evolution can only permit us to calculate all the force components acting upon charges. The magnetic field is merely a spatial derivative of the vector field.

As follows from the above consideration, it is more appropriate to write the Lotentz force in terms of the magnetic vector potential

$$\vec{F}' = e\,\vec{E} + e\,[\vec{V} \times rot\,\vec{A}_B] = e\,\vec{E} - e(\vec{V}\nabla)\vec{A}_B + e\,grad\,(\vec{V}\,\vec{A}_B), \quad (1.12)$$

which visualizes the complete structure of the force.

The Faraday law, Eq. (1.2) is referred to as the law of electromagnetic induction because it shows how varying magnetic fields can generate electric fields. However, classical electrodynamics contains no law of magnetoelectric induction showing how magnetic fields can be excited by varying electric fields. This aspect of classical electrodynamics evolved along a different pathway. First, the law

$$\oint \vec{H}\,d\vec{l} = I, \quad (1.13)$$

was known, in which $I$ was the current crossing the area of the integration contour. In the differential from Eq. (1.13) becomes

$$rot\,\vec{H} = \vec{j}_\sigma, \quad (1.14)$$

where $\vec{j}_\sigma$ is the conduction current density.

Maxwell supplemented Eq. (1.14) with displacement current

$$rot\,\vec{H} = \vec{j}_\sigma + \dfrac{\partial \vec{D}}{\partial t}. \quad (1.15)$$

However, if Faraday had performed measurement in varying electric induction fluxes, he would have inferred the following law

$$\oint \vec{H}'d\vec{l}' = \dfrac{d\Phi_D}{dt}, \quad (1.16)$$

where $\Phi_D = \int \vec{D}\,d\vec{S}'$ is the electric induction flux. Then

$$\oint \vec{H}'d\vec{l}' = \int \dfrac{\partial \vec{D}}{\partial t}d\vec{S} + \oint [\vec{D} \times \vec{V}]d\vec{l}' + \int \vec{V}\,div\,\vec{D}\,d\vec{S}'. \quad (1.17)$$

Unlike $div \vec{B} = 0$ in magnetic fields, electric fields are characterized by $div \vec{D} = \rho$ and the last term in the right-hand side of Eq. (1.17) describes the conduction current *I*, i.e. the Ampere law follows from Eq. (1.16). Eq. (1.17) gives

$$\vec{H} = [\vec{D} \times \vec{V}], \quad (1.18)$$

which was earlier obtainable only from the Lorentz transformation.

Moreover, as was shown convincingly in [2], Eq. (1.18) also leads out of the Biot-Savart law if magnetic fields are calculated from the electric fields excited by moving charges. In this case the last term in the right-hand side of Eq. (1.17) can be omitted and the induction laws become completely symmetrical.

$$\oint \vec{E}' d\vec{l}' = -\int \frac{\partial \vec{B}}{\partial t} dS - \oint [\vec{B} \times \vec{V}] d\vec{l}',$$

$$\oint \vec{H}' d\vec{l}' = \int \frac{\partial \vec{D}}{\partial t} dS + \oint [\vec{D} \times \vec{V}] d\vec{l}'. \quad (1.19)$$

$$E' = \vec{E} + [\vec{V} \times \vec{B}],$$
$$H' = \vec{H} - [\vec{V} \times \vec{D}]. \quad (1.20)$$

Earlier, Eqs. (1.20) were only obtainable from the covariant Lorentz transformations, i.e. in the framework of special theory of relativity (STR). Thus, the STR results accurate to the $\sim \frac{V}{c}$ terms can be derived from the induction laws through the Galilean transformations. The STR results accurate to the $\frac{V^2}{c^2}$ terms can be obtained through transformation of Eq (1.19). At first, however, we shall introduce another vector potential which is not used in classical electrodynamics. Let us assume for vortex fields [3] that

$$\vec{D} = rot \vec{A}_D, \quad (1.21)$$

where $\vec{A}_D$ is the electric vector potential. It then follows from Eq. (1.19) that

$$\vec{H}' = \frac{\partial \vec{A}_D}{\partial t} + [\vec{V} \nabla] \vec{A}_D - grad [\vec{V} \vec{A}_D], \quad (1.22)$$

or

$$\vec{H}' = \frac{\partial \vec{A}_D}{\partial t} - [\vec{V} \times rot \vec{A}_D], \quad (1.23)$$

or

$$\vec{H}' = \frac{d \vec{A}_D}{dt} - grad [\vec{V} \vec{A}_D]. \quad (1.24)$$

These equations present the law of magnetoelectric induction written in terms of the electric vector potential.

To illustrate the importance of the introduction of the electric vector potential, we come back to an infinitely long solenoid. The situation is much the same, and the only change is that the vectors $\vec{B}$ are replaced with the vectors $\vec{D}$. Such situation is quite realistic: it occurs when the space between the flat capacitor plates is filled with high electric inductivities. In this case the displacement flux is almost entirely inside the dielectric. The attempt to calculate the magnetic field outside the space occupied by the dielectric (where $\vec{D} \cong 0$) runs into the same problem that existed for the calculation beyond the fields $\vec{E}$ of an infinitely long solenoid. The introduction of the electric vector potential permits a correct solution of this problem. This however brings up the question of priority: what is primary and what is secondary? The electric vector potential is no doubt primary because electric vortex fields are excited only where the rotor of such potential is non-zero.

As follows from Eqs. (1.20), if the reference systems move relative to each other, the fields $\vec{E}$ and $\vec{H}$ are mutually connected, i.e. the movement in the fields $\vec{H}$ induces the fields $\vec{E}$ and vice versa. But new consequences appear, which were not considered in classical electrodynamics. For illustration, let us analyze two parallel conducting plates with the electric field $\vec{E}$ in between. In this case the surface charge $\rho_S$ per unit area of each plate is $\varepsilon E$. If the other reference system is made to move parallel to the plates in the field $E$ at the velocity $\Delta V$, this motion will generate an additional field $\Delta H = \Delta V \varepsilon E$. If a third reference system starts to move at the velocity $\Delta V$, within the above moving system, this motion in the field $\Delta H$ will generate $\Delta E = \mu \varepsilon \Delta V^2 E$, which is another contribution to the field $E$. The field $E'$ thus becomes stronger in the moving system than it is in the stationary one. It is reasonable to suppose that the surface charge at the plates of the initial system has increased by $\mu \varepsilon^2 \Delta V^2 E$ as well.

If we put $\vec{E}_{||}$ and $\vec{H}_{||}$ for the field components parallel to the velocity direction and $\vec{E}_\perp$ and $\vec{H}_\perp$ for the perpendicular components, the final fields at the velocity $V$ can be written as

$$\vec{E}'_{||} = \vec{E}_{||},$$

$$\vec{E}'_\perp = \vec{E}_\perp c h \frac{V}{c} + \frac{Z_0}{V}[\vec{V} \times \vec{H}_\perp] s h \frac{V}{c},$$

$$\vec{H}'_{||} = \vec{H}_{||}, \qquad (1.25)$$

$$\vec{H}'_\perp = \vec{H}_\perp c h \frac{V}{c} - \frac{1}{Z_0 V}[\vec{V} \times \vec{E}_\perp] s h \frac{V}{c},$$

where $Z_0 = \sqrt{\dfrac{\mu}{\varepsilon}}$ is the space impedance, $c = \sqrt{\dfrac{1}{\mu\varepsilon}}$ is the velocity of light in the medium under consideration.

The results of these transformations coincide with the STR data with the accuracy to the $\sim \dfrac{V^2}{c^2}$ terms. The higher-order corrections do not coincide. It should be noted that until now experimental tests of the special theory of relativity have not gone beyond the $\sim \dfrac{V^2}{c^2}$ accuracy.

As an example, let us analyze how Eqs. (1.25) can account for the phenomenon of phase aberration which was inexplicable in classical electrodynamics.

Assume that there are plane wave components $H_Z$ and $E_X$, and the primed system is moving along the $x$-axis at the velocity $V_X$. The field components with in the primed coordinates can be written as

$$E_X' = E_X,$$
$$E_Y' = H_Z sh\dfrac{V_x}{c},$$
$$H_Z' = H_Z ch\dfrac{V_X}{c}.$$
(1.27)

The total field $E$ in the moving system is

$$E' = \left[\left(E_X'\right)^2 + \left(E_Y'\right)^2\right]^{1/2} = E_X\, ch\,\dfrac{V_X}{c}.$$
(1.28)

Hence, the Poynting vector no longer follows the direction of the $y$-axis. It is in the $xy$-plane and tilted about the $y$-axis at an angle determined by Eqs. (1.27). The ratio between the absolute values of the vectors $E$ and $H$ is the same in both the systems. This is just what is known as phase aberration in classical electrodynamics.

## 2. The force interaction between current-carreging conductors

As follows from the transformations in Eq. (1.25) if two charges move at the relative velocity $\vec{V}$, their interaction is determined not only by the absolute values of the charges but by the relative motion velocity as well. The new value of the interaction force is found as

$$\vec{F} = \dfrac{g_1 g_2\, ch\dfrac{V_\perp}{c}}{4\pi\,\varepsilon} \cdot \dfrac{\vec{r}_{12}}{r_{12}^3},$$
(2.1)

where $\vec{r}_{12}$ is the vector connecting the charges, $V_\perp$ is the component of the velocity $\vec{V}$, normal to the vector $\vec{r}_{12}$.

If opposite-sign charges are engaged in the relative motion, their attraction increases. If the charges have the same signs, their repulsion enhances. For $\vec{V} = 0$, Eq. (2.1) becomes the Coulomb law.

Using Eq. (2.1), a mew value of the potential $\varphi(r)$ can be introduced at the point, where the charge $g_2$ is located, assuming that $g_2$ is immobile and only $g_1$ executes the relative motion

$$\varphi(r) = \frac{g_1 \, ch \frac{V_\perp}{c}}{4\pi \, \varepsilon \, r} . \qquad (2.2)$$

We can denote this potential as "scalar-vector", because its value is dependent not only on the charge involved but on the value and the direction of its velocity as well. The potential energy of the charge interaction is

$$W = \frac{g_1 \, g_2 \, ch \frac{V_\perp}{c}}{4\pi \, \varepsilon \, r} . \qquad (2.3)$$

Eqs. (2.1), (2.2) and (2.3) apparently account for the change in the value of the moving charges.

Using these equations, it is possible to calculate the force of the conductor-current interactions and allow, through superposition, for the interaction forces of all moving and immobile charges in the conductors. We thus obtain all currently existing laws of electromagneticm.

Let us examine the force, interaction of two $z$-spaced conductors (Fig. 2.1) assuming that the electron velocities in the conductors are $V_1$ and $V_2$. The moving charge values per unit length of the conductors are $g_1$ and $g_2$.

In terms of the present-day theory of electromagnetism, the forces of the interaction of the conductors can be found by two methods.

One of the conductors (e.g., the lower one) generates the magnetic field $H(r)$ in the location of the first conductor. This field is

$$H(r) = \frac{g_1 V_1}{2\pi \, r} . \qquad (2.4)$$

The field $E'$ is excited in the coordinate system moving together with the charges of the upper conductor:

$$E' = [\vec{V} \times \vec{B}] = V_2 \, \mu \, H(r) . \qquad (2.5)$$

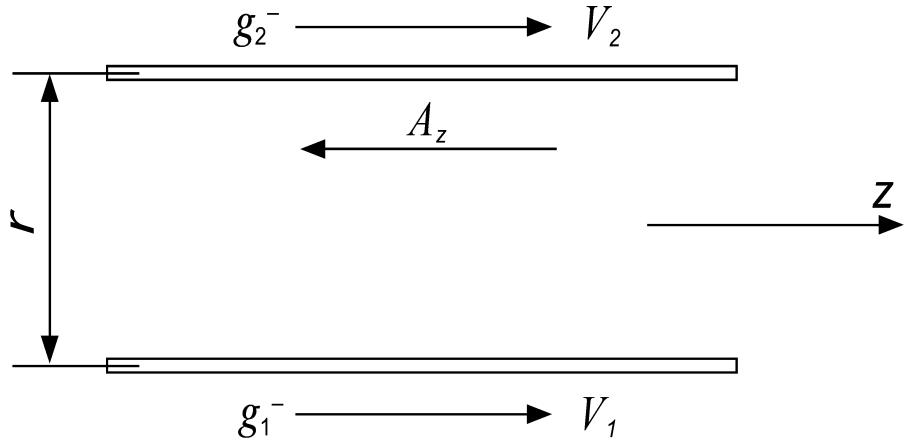

Fig. 2.1. Schematic view of force interaction between current-carreging conductors of a two-conductor line in terms of the present-day model.

I.e. the charges moving in the upper conductor experience the Lorentz force. This force per unit length of the conductor is

$$F = \frac{\mu \, g_1 V_1 \, g_2 V_2}{2\pi \, r} = \frac{I_1 I_2}{2\pi \, \varepsilon \, c^2 \, r} \,. \tag{2.6}$$

Eq. (2.6) can be obtained in a different way. Assume that the lower conductor excites a vector potential in the region of the upper conductor. The $z$–component of the vector potential is

$$A_z = -\frac{g_1 V_1 \ln r}{2\pi \, \varepsilon \, c^2} = -\frac{I_1 \ln r}{2\pi \, \varepsilon \, c^2} \,. \tag{2.7}$$

The potential energy per unit length of the upper conductor carrying the current $I_2$ in the field of the vector potential $A_z$ is

$$W = I_2 A_z = -\frac{I_1 I_2 \ln r}{2\pi \, \varepsilon \, c^2} \,. \tag{2.8}$$

Since the force is the derivative of the potential energy with respect to the opposite-sign coordinate, it is written as

$$F = -\frac{\partial W}{\partial r} = \frac{I_1 I_2}{2\pi \, \varepsilon \, c^2 r} \,. \tag{2.9}$$

Both the approaches show that the interaction force of two conductors is the result of the interaction of moving charges: some of them excite fields, the others interact with them. The immobile charges representing the lattice do not participate in the interaction in this scheme. But the forces of the magnetic interaction between the conductors act just on the lattice. Classical electrodynamics does mot explain how the moving charges experiencing this force can transfer it to the lattice.

The above models of iteration are in unsolvable conflict, and experts in classical electrodynamics prefer to pass it over in silence. The conflict is connected with estimation of the interaction force of two parallel-moving charges. Within the above models such two charges should be attracted. Indeed, the induction $B$ caused by the moving charge $g_1$ at the distance $r$ is

$$B = \frac{g_1 V}{2\pi \, \varepsilon \, c^2 r^2} \; . \tag{2.10}$$

If another charge $g_2$ moves at the same velocity $V$ in the same direction at the distance $r$ from the first charge, the induction $B$ at the location of $g_2$ produces the force attracting $g_1$ and $g_2$.

$$F = \frac{g_1 g_2 V^2}{4\pi \, \varepsilon \, c^2 r^2} . \tag{2.11}$$

An immovable observer would expect these charges to experience attraction along with the Coulomb repulsion. For an observer moving together with the charges there is only the Coulomb repulsion and no attraction. Neither classical electrodynamics not the special theory of relativity can solve the problem.

Physically, the introduction of magnetic fields reflects certain experimental facts, but so far we can hardly understand where these fields come from.

It is useful to analyze here the interaction of current-carrying systems in terms of Eqs. (2.1), (2.2) and (2.3) [4, 5].

We come back again to the interaction of two thin conductors with charges moving at the velocities $V_1$ and $V_2$ (Fig. 2.2).

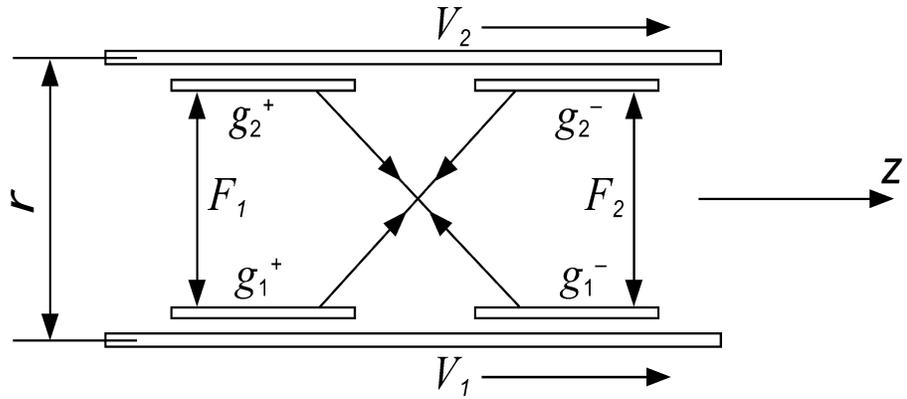

Fig. 2.2. Schematic view of force interaction between current-carrying wires of a two-conductor line. The lattice is charged positively.

$g_1^+$, $g_2^+$ and $g_1^-$, $g_2^-$ are the immobile and moving charges, respectively, pre unit length of the conductors. $g_1^+$ and $g_2^+$ refer to the positively charged lattice in the lower and upper conductors, respectively. Before the charges start moving, both the conductors are assumed to be neutral electrically, i.e. they contain the same number of positive and negative charges.

Each conductor has two systems of unlike charges with the specific densities $g_1^+$, $g_1^-$ and $g_2^+$, $g_2^-$. The charges neutralize each other electrically. To make the analysis of the interaction forces more convenient, in Fig. 2.2 the systems are separated along the z-axis. The negative-sign subsystems (electrons) have velocities $V_1$ and $V_2$. The force of the interaction between the lower and upper conductors can be considered as a sum of four forces specified in Fig. 2.2 (the direction is shown by arrows). The attraction forces $F_3$ and $F_4$ are positive, and the repulsion forces $F_1$ and $F_2$ are negative.

According to Eq. (2.1), the forces between the individual charge subsystems (Fig. 2.2) are

$$F_1 = -\frac{g_1^+ g_2^+}{2\pi \varepsilon r},$$

$$F_2 = -\frac{g_1^- g_2^-}{2\pi \varepsilon r} ch\frac{V_1 - V_2}{c},$$

$$F_3 = +\frac{g_1^- g_2^+}{2\pi \varepsilon r} ch\frac{V_1}{c}, \qquad (2.12)$$

$$F_4 = +\frac{g_1^+ g_2^-}{2\pi \varepsilon r} ch\frac{V_2}{c}.$$

By adding up the four forces and remembering that the product of unlike charges and the product of like charges correspond to the attraction and repulsion forces, respectively, we obtain the total specific force per unit length of the conductor

$$F_\Sigma = \frac{g_1 g_2}{2\pi \varepsilon r}\left(ch\frac{V_1}{c} + ch\frac{V_2}{c} - ch\frac{V_1 - V_2}{c} - 1\right). \qquad (2.13)$$

where $g_1$ and $g_2$ are the absolute values of charges. The signs of the forces appear in the bracketed expression. Assuming $V \ll c$, we use only the two first terms in the expression of $ch\frac{V}{c}$, i.e. $ch\frac{V}{c} \cong 1 + \frac{1}{2}\frac{V^2}{c^2}$. Eq. (2.13) gives

$$F_{\Sigma 1} = \frac{g_1 V_1 g_2 V_2}{2\pi \varepsilon c^2 r} = \frac{I_1 I_2}{2\pi \varepsilon c^2 r}, \qquad (2.14)$$

where $g_1$ and $g_2$ are the absolute values of specific charges, and $V_1$, $V_2$ are taken with their signs.

It is seen that Eqs. (2.6), (2.9) and (2.13) coincide though they were obtained by different methods.

According to Feynman [2], the e.m.f. of the circuit can be interpreted using two absolutely different laws. The paradox has however been clarified. The force of the enteraction between the current-carrying systems can be obtained even by three absolutely different methods. But in the third method, the motion "magnetic field" is no longer necessary and the lattice can directly participate in the formation of the interaction forces. This was impossible with the previous two techniques.

In practice the third method however runs into a serious obstacle. Assuming $g_2^+ = 0$ and $V_2 = 0$, i.e. the interaction, for example, between the lower current-carrying system and the immobile charge $g_2^-$ the interaction force is

$$F_{\Sigma 2} = -\frac{1}{2} \cdot \frac{g_1 g_2 V_1^2}{2\pi \varepsilon c^2 r} . \qquad (2.15)$$

This means that the current in the conductor is not electrically neutral, and the electric field

$$E_\perp = \frac{g_1 V_1^2}{4\pi \varepsilon c^2 r}, \qquad (2.16)$$

is excited around the conductor, which is equivalent to an extra specific static charge on the conductor

$$g = -g_1 \frac{V_1^2}{c^2} . \qquad (2.17)$$

When Faraday and Maxwell formulated the basic laws of electrodynamics, it was impossible to confirm Eq. (3.17) experimentally because the current densities in ordinary conductors are too small to detect the effect. The assumption that the charge is independent of its velocity and the subsequent introduction of a magnetic field were merely voluntaristic acts.

In superconductors the current densities permit us to find the correction for the charge $\sim g\frac{V_1^2}{c^2}$ experimentally. Initially, [6] was taken as evidence for the dependence of the value of the charge on its velocity. The author of this study has also investigated this problem [4,5], but, unlike [6], in his experiments current was introduced into a superconducting coil by an inductive non-contact method. Even in this case a charge appeared on the coil [4,5]. The experimental objects were superconducting composite Nb – Ti wires coated with copper, and it is not cleat what mechanism is responsible for the charge on the coil. It may be brought by mechanical deformation which causes a displacement of the Fermi level in the copper. Experiments on non-coated superconducting wires may be more informative. Anyhow, the subject has not been exhausted and further experimental findings are of paramount importance to fundamental physics. Using this model, we should remember that there is no reliable experimental data on static electric fields around the conductor. According to Eq. (2.16), such fields are excited because the value of the charge is dependent on its velocity. Is there any physical mechanism which could maintain the interacting current-carrying systems electrically neutral within this model? Such mechanism does exist. To explain it, let us consider the current-carrying circuit in Fig. 2.3. This is a superconducting thin film whose thickness is smaller than the field penetration depth in the superconductor. The current is therefore distributed uniformly over the film thickness. Assume that the bridge connect-ing the wide parts of the film is much narrower than the rest of the current-carrying film. If persistent current is excited in such a circuit, the current density and hence the current carrier velocity $V_1$ in the bridge will much exceed the velocity $V_0$ in the wide parts of the film.

Such situation is possible if the current carriers are accelerated in the part $d_1$ and slowed down in the part $d_2$. But acceleration and slowing-down of charges is possible only in electric fields. If $V_1 > V_0$, the potential difference between the parts $d_1$ and $d_2$ which causes acceleration or slowing-down is determined as

$$U = \frac{m V_1^2}{2 e} \quad . \tag{2.18}$$

This potential difference can appear only due to the charge density gradient in the parts $d_1$ and $d_2$, i.e. the density of charge carriers decreases with acceleration and increases with slowing down. The relation $n_0 > n_1$ should be fulfilled, where $n_0$ and $n_1$ are the current-carrier densities in the wide and narrow bridge parts of the film, respectively. It is clear that some energy is needed to accelerate charges which have masses. Let us find out where this energy comes from.

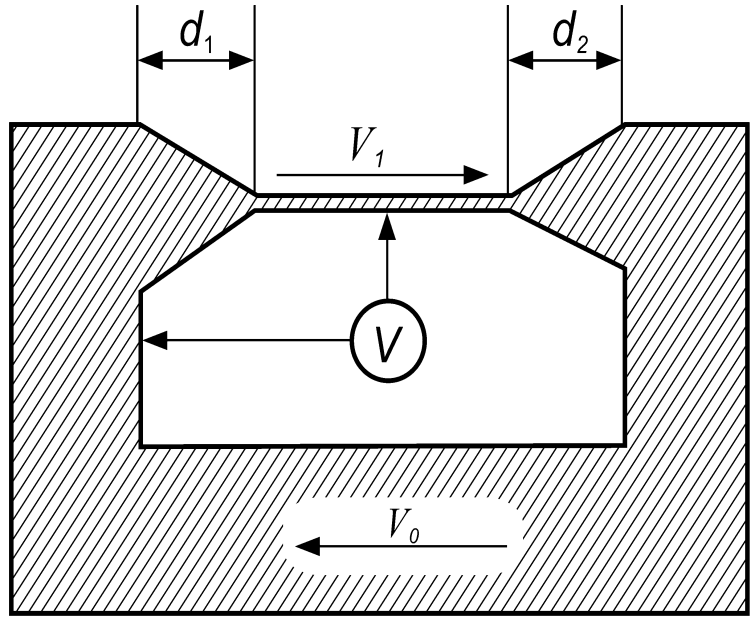

Fig.2.3.   Schematic view of a current-carrying circuit based on a supercon ducting film.

On acceleration the electrostatic energy available in the electrostatic field of the current carriers converts into kinetic energy. The difference in electrostatic energy between two identical volumes having different electron densities can be written as

$$\Delta W = \Delta n \frac{e^2}{8\pi \varepsilon r}, \qquad (2.19)$$

where $\Delta n = n_0 - n_1$, $e$ is the electron charge, $r$ is the electron radius.
Since

$$\frac{e^2}{8\pi \varepsilon r} = m c^2, \qquad (2.20)$$

where $m$ is the electron mass, Eq. (2.19) can be rewritten as

$$\Delta W = \Delta n\, m\, c^2. \qquad (2.21)$$

This energy is used to accelerate the current carriers.
Hence,

$$\Delta W = \frac{n_0\, m\, V_1^2}{2}, \qquad (2.22)$$

and

$$\Delta n = n_0 \frac{1}{2} \cdot \frac{V_1^2}{c^2}. \qquad (2.23)$$

It is shown that the concentration of electrons in an electron beam moving throughan electrically neutral medium is less than that in an immovable beam an is defined by the relation.

$$n_1 = n_0 \left(1 - \frac{1}{2} \cdot \frac{V_1^2}{c^2}\right). \qquad (2.24)$$

In this case, the contradiction mentioned above is absent.

Certainly, the given approach is nonconventional, but it is very evident, since it is clearly clear, what forces operate between the charged particles which are carrying out mutual movement. Clearly, that the considered forces will lead to compression a cord of the current, and forces of compression will be more, than forces of pushing apart.

We should note one more circumstance. That density of the electrons which move through a lattice, less, than density motionless of the electrons, specified still F.London [7], however, for electrodynamics and this circumstance of greater consequences had no thermodynamics of superconductors, since such amendments are very small.

However, for plasma, especially in case of pinch-effect when density of currents can reach greater sizes, this question gets special value. At greater density of current the difference of longitudinal density of the electrons and ions can reach greater sizes, and it will lead to presence of greater longitudinal electric fields. These fields will try to break off a plasma cord in longitudinal a direction, leading a additional instability about which we earlier did not know. Thus, the new approach opens a way not only to the best physical understanding of pinch-effect, but also predicts the new phenomena which, certainly, it is necessary to consider at research and practical use of pinch-effect.

We see that the change in the current-carrier density is quite small, but this change is just responsible for the existence of the longitudinal electric field accelerating or slowing down the charges in the parts $d_1$ and $d_2$. Let us call such fields "configuration fields" as they are connected with a certain configuration of the conductor. These fields are available in normal conductors too, but they are much smaller than the fields related to the Ohmic resistance.

We can expect that a voltameter connected to the circuit, like is shown in Fig. 2.3, would be capable of registering the configuration potential difference in accordance with Eq. (2.18). If we used an ordinary liquid and a manometer instead of a voltameter, according to the Bernoulli equation, the manometer could register the pressure difference. For lead films, the configuration potential difference is ~$10^{-7}$ B, though it is not observablt experimentally. We can explain this before hand. As the velocities of the current carriers increase and their densities decrease, the electric fields njrmal to their motion enhance. These two precesses counterbalance each other. As a result, the normal component of the electric field has a zero balue in all parts of the film. In terms of the considered, this looks like

$$F_1 = -\frac{g_1^+ g_2^+}{2\pi \, \varepsilon \, r},$$

$$F_2 = -\frac{g_1^- g_2^-}{2\pi \, \varepsilon \, r}\left(1 - \frac{1}{2}\cdot\frac{V_1^2}{c^2}\right)\cdot\left(1 - \frac{1}{2}\cdot\frac{V_2^2}{c^2}\right) ch\frac{V_1 - V_2}{c},$$

$$F_3 = \frac{g_1^- g_2^+}{2\pi \, \varepsilon \, r}\left(1 - \frac{1}{2}\cdot\frac{V_1^2}{c^2}\right) ch\frac{V_1}{c}, \qquad (2.25)$$

$$F_4 = \frac{g_1^+ g_2^-}{2\pi \, \varepsilon \, r}\left(1 - \frac{1}{2}\cdot\frac{V_2^2}{c^2}\right) ch\frac{V_1}{c}.$$

The bracketed expressions in Eqs. (2.25) allow for the motion-related change in the density of the charges $g_1^-$ and $g_2^-$.

After expanding $ch$, multiplying out and allowing only for the ~ $V^2/c^2$ terms, Eqs. (2.25) give

$$F_1 \cong -\frac{g_1^+ g_2^+}{2\pi \varepsilon r},$$

$$F_2 \cong -\frac{g_1^- g_2^-}{2\pi \varepsilon r}\left(1 - \frac{V_1 V_2}{c^2}\right),$$

$$F_3 \cong \frac{g_1^- g_2^+}{2\pi \varepsilon r}, \qquad (2.26)$$

$$F_4 \cong \frac{g_1^+ g_2^-}{2\pi \varepsilon r}.$$

By adding up $F_1$, $F_2$, $F_3$ and $F_4$, we obtain the total force of the interaction

$$F_\Sigma = \frac{g_1^- V_1 \, g_2^- V_2}{2\pi \varepsilon c^2 r} = \frac{I_1 I_2}{2\pi \varepsilon c^2 r}. \qquad (2.27)$$

Again, we have a relation coinciding with Eqs. (2.6) and (2.9). However, in this case the current-carrying conductors are neutral electrically. Indeed, if we analyze the force interaction. For example, between the lower conductor and the upper immobile charge $g_2$ (putting $g_2^+ = 0$ and $V_2 = 0$), the total interaction force will be zero, i.e. the conductor with flowing current is electrically neutral.

If we consider the interaction of two parallel – moving electron flows (taking $g_1^+ = g_2^+ = 0$ and $V_1 = V_2$), according to Eq. (2.12), the total force is

$$F_\Sigma = -\frac{g_1^- g_2^-}{2\pi \varepsilon r}. \qquad (2.28)$$

It is seen that two electron flows moving at the same velocity in the absence of a lattice experience only the Coulomb repulsion and no attraction included into the magnetic field concept.

Physically, in this model the force interaction of the current-carrying systems is not connected with any now field. The interaction is due to the enhancement of the electric fields normal to the direction of the charge motion.

The phenomenological concept of the magnetic field of correct only when the charges of the current carriers are compensated with the charges of the immobile lattice, the current carriers excite a magnetic field. The magnetic field concept is not correct for freely moving charges when there are no compensating charges of the lattice. In this case a moving charged particle or a flow of charged particles does not excite a magnetic field. Thus, the concept of the phenomenological magnetic field is true but for the above case.

It is easy to show that using the scalar-vector potential, we can obtain all the presently existing laws of magnetism. Besides, the approach proposed permits a solution of the problem of the interaction between two parallel-moving charges which could not be solved in terms of the magnetic field concept.

# 3. Retarded potentials

Whatever occurs in electrodynamic, it is connected with the interaction of moving and immobile charges. The introduction of the scalar-vector potential answers this question. The potential is based on the laws of electromagnetic and magnetoelectric induction. The Maxwell equations describing the wave processes in material media also follow from these laws. The Maxwell equations suggest that the velocity of field propagation is finite and equal to the velocity of light.

The problem of electromagnetic radiation can be solved of the elementary level using the scalar-vector potential and the finiteness of propagation of electric processes.

For this purpose, the retarded scalar-vector potential

$$\varphi(r',t) = \frac{g_1 \, ch\frac{V'_\perp}{c}}{4\pi \, \varepsilon \, r'}, \qquad (3.1)$$

is introduced, where $V'_\perp$ is the velocity of the charge $g_1$ at the moment $t' = t - \frac{r'}{c}$, normal to the vector $\vec{r}'$, $r'$ is the distance between the charge $g_1$ and point 2, where the field is sought for at the moment $t$. The field at point 2 can be found from the relation $\vec{E} = -\operatorname{grad}\varphi$. Assume that at the moment $t - \frac{r'}{c}$ the charge $g_1$ is at the origin of the coordinates and its velocity is $V'_\perp(t)$ (Fig. 3.1). The field $E_y$ at point 2 is

$$E_y = -\frac{\partial \varphi(2,t)}{\partial y} = -\frac{e_0}{4\pi \, \varepsilon \, r'} \cdot \frac{\partial}{\partial y} ch\frac{V'_\perp(t)}{c}. \qquad (3.2)$$

Differentiation is performed assuming $r'$ to be a constant magnitude. From Eq. (3.2) we obtain

$$E_y = -\frac{e_0}{4\pi \, \varepsilon \, c \, r'} \cdot \frac{\partial V'_\perp(t)}{\partial y} sh\frac{V'_\perp(t)}{c} = -\frac{e_0}{4\pi \, \varepsilon \, c \, r'} \cdot \frac{1}{V'_\perp(t)} \cdot \frac{\partial V'_\perp(t)}{\partial t} sh\frac{V'_\perp(t)}{c}. \qquad (3.3)$$

Using only the first term of the expansion of $sh\frac{V'_\perp(t)}{c}$ we can obtain from Eq. (3.3)

$$E_y = -\frac{e_0}{4\pi \, \varepsilon \, c \, r'} \cdot \frac{\partial V'_\perp(t)}{\partial t}. \qquad (3.4)$$

This law of radiation from a moving charge is well known though its derivation is more complex [2].

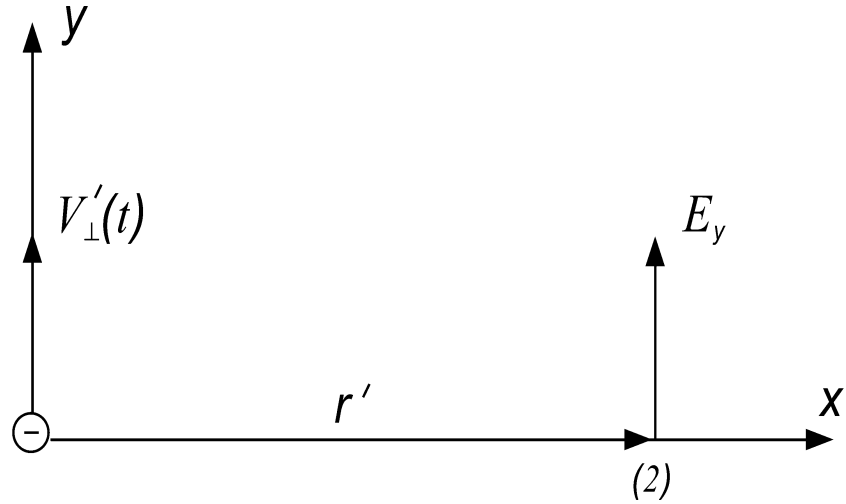

Fig. 3.1. Formation of the retarded scalar-vector potential.

All the problems of radiation can be solved at the elementary level using Eq. (3.4) . this equation is also the induction law assuming that the retardation time is very short.

Now we should write all equations that are necessary for solving existing problems of electrodynamics. Relation (3.1) must be supplemented by the wave equation for electric fields. To derive the latter, let us write the Maxwell equations in the form already known for us

$$rot\,\vec{E} = -\mu^*(\omega)\frac{\partial \vec{H}}{\partial t}$$
$$rot\,\vec{H} = \varepsilon^*(\omega)\frac{\partial \vec{E}}{\partial t}$$
(3.5)

Expressions $\mu^*(\omega)$ и $\varepsilon^*(\omega)$ are some frequency-dependent functionals that relate to each other the frequency and spatial derivatives of electric and magnetic fields. Eliminating magnetic field from relations (3.5) we arrive at the wave equation

$$rot\,rot\,\vec{E} + \mu^*(\omega)\,\varepsilon^*(\omega)\frac{\partial^2 \vec{E}}{\partial t^2} = 0 \qquad (3.6)$$

Relation (3.6) indicates that electric fields propagate through space at a certain velocity, therefore, we can introduce delayed potentials in accordance with relation (3.1).

It is surprising that Eq. (3.1) actually accounts for the whole of electrodynamics beause all current electrodynamics problems can be solved using this equation. What is then a magnetic field? This is merely a convenient mathematical procedure which is not necessarily gives a correct result (e.g., in the case of parallel-moving charges). Now we can state that electrocurrent, rather than electromagnetic, waves travel in space. Their electric field and displacement current vectors are in the same plane and displaced by $\pi/2$.

In terms of Eq. (3.1), electrodynamics and optics can be reconstructed completely to become simpler, more intelligible and obvious.

## 4. Dispersion of dielectric and magnetic permittivities of physical media

It is well known that electric and magnetic inductivities of material media can depend on frequency, i.e. they can exhibit dispersion. But even Maxwell himself, who was the author of the basic equations of electrodynamics, believed that $\varepsilon$ and $\mu$ were frequency-independent fundamental constants.

How the idea of $\varepsilon$ and $\mu$-dispersion appeared and evolved is illustrated vividly in the monograph of well-known specialists in physics of plasma [8]: while working at the equations of electrodynamics of material, media, G. Maxwell looked upon electric and magnetic inductivities as constants (that is why this approach was so lasting). Much later, at the beginning of the XX century, G. Heavisidr and R.Wull put forward their explanation for phenomena of optical dispersion (in particular rainbow) in which electric and magnetic inductivities came as functions of frequency. Quite recently, in the mid-50ies of the last century, physicists arrived at the conclusion that these parameters were dependent not only on the frequency but on the wave vector as well. That was a revolutionary breakaway from the current concepts. The importance of the problem is clearly illustrated by what happened at a seminar held by L. D. Landau in 1954, where he interrupted A. L. Akhiezer reporting on the subject: "Nonsense, the refractive index cannot be a function of the refractive index". Note, this was said by L. D. Landau, an outstanding physicist of our time.

What is the actual situation? Running ahead, I can admit that Maxwell was right: both $\varepsilon$ and $\mu$ are frequency – independent constants characterizing one or another material medium. Since dispersion of electric and magnetic inductivities of material media is one of the basic problems of the present – day physics and electrodynamics, the system of views on these questions has to be radically altered again (for the second time!).

In this context the challenge of this study was to provide a comprehensive answer to the above questions and thus to arrive at a unified and unambiguous standpoint. This will certainly require a revision of the relevant interpretations in many fundamental works.

## 4.1. Plasma media

It is noted that dispersion of electric and magnetic inductivities of material media is a commonly accepted idea. The idea is however not correct.

To explain this statement and to gain a better understanding of the physical essence of the problem, we start with a simple example showing how electric lumped-parameter circuits can be described. As we can see below, this example is directly concerned with the problem of our interest and will give us a better insight into the physical picture of the electrodynamic processes in material media.

In a parallel resonance circuit including a capacitor $C$ and an inductance coil $L$, the applied voltage $U$ and the total current $I_\Sigma$ through the circuit are related as

$$I_\Sigma = I_C + I_L = C\frac{dU}{dt} + \frac{1}{L}\int U\,dt, \qquad (4.1)$$

where $I_C = C\dfrac{dU}{dt}$ is the current through the capacitor, $I_L = \dfrac{1}{L}\int U\,dt$ is the current through the inductance coil. For the harmonic voltage $U = U_0 \sin \omega t$

$$I_\Sigma = \left(\omega C - \frac{1}{\omega L}\right) U_0 \cos \omega t. \qquad (4.2)$$

The term in brackets is the total susceptance $\sigma_x$ of the circuit, which consists of the capacitive $\sigma_c$ and inductive $\sigma_L$ components

$$\sigma_x = \sigma_c + \sigma_L = \omega C - \frac{1}{\omega L}. \qquad (4.3)$$

Eq. (4.2) can be re-written as

$$I_\Sigma = \omega C\left(1 - \frac{\omega_0^2}{\omega^2}\right) U_0 \cos \omega t, \qquad (4.4)$$

where $\omega_0^2 = \dfrac{1}{LC}$ is the resonance frequency of a parallel circuit.

From the mathematical (i.e. other than physical) standpoint, we may assume a circuit that has only a capacitor and no inductance coil. Its frequency – dependent capacitance is

$$C^*(\omega) = C\left(1 - \frac{\omega_0^2}{\omega}\right). \qquad (4.5)$$

Another approach is possible, which is correct too.
Eq. (4.2) can be re-written as

$$I_\Sigma = -\frac{\left(\dfrac{\omega^2}{\omega_0^2}-1\right)}{\omega L} U_0 \cos \omega t \ . \tag{4.6}$$

In this case the circuit is assumed to include only an inductance coil and no capacitor. Its frequency – dependent inductance is

$$L^*(\omega) = \frac{L}{\left(\dfrac{\omega^2}{\omega_0^2}-1\right)} \ . \tag{4.7}$$

Using the notion of Eqs. (4.5) and (4.7), we can write

$$I_\Sigma = \omega \, C^*(\omega) U_0 \cos \omega t \ , \tag{4.8}$$

or

$$I_\Sigma = -\frac{1}{\omega L^*(\omega)} U_0 \cos \omega t \ . \tag{4.9}$$

Eqs (4.8) and (4.9) are equivalent and each of them provides a complete mathematical description of the circuit. From the physical point of view, $C^*(\omega)$ and $L^*(\omega)$ do not represent capacitance and inductance though they have the corresponding dimensions. Their physical sense is as follows:

$$C^*(\omega) = \frac{\sigma_X}{\omega} \ , \tag{4.10}$$

i.e. $C^*(\omega)$ is the total susceptance of this circuit divided by frequency:

$$L^*(\omega) = \frac{1}{\omega \, \sigma_X} \ , \tag{4.11}$$

and $L^*(\omega)$ is the inverse value of the product of the total susceptance and the frequency.

Amount $C^*(\omega)$ is constricted mathematically so that it includes $C$ and $L$ simultaneously. The same is true for $L^*(\omega)$.

We shall not consider here any other cases, e.g., series or more complex circuits. It is however important to note that applying the above method, any circuit consisting of the reactive components $C$ and $L$ can be described either through frequency – dependent inductance or frequency – dependent capacitance.

But this is only a mathematical description of real circuits with constant – value reactive elements.

It is well known that the energy stored in the capacitor and inductance coil can be found as

$$W_C = \frac{1}{2} C U^2 \ , \tag{4.12}$$

$$W_L = \frac{1}{2} L I^2 . \tag{4.13}$$

But what can be done if we have $C^*(\omega)$ and $L^*(\omega)$? There is no way of substituting them into Eqs. (4.12) and (4.13) because they can be both positive and negative. It can be shown readily that the energy stored in the circuit analyzed is

$$W_\Sigma = \frac{1}{2} \cdot \frac{d\,\sigma_X}{d\,\omega} U^2 , \tag{4.14}$$

or

$$W_\Sigma = \frac{1}{2} \cdot \frac{d[\omega\, C^*(\omega)]}{d\,\omega} U^2 , \tag{4.15}$$

or

$$W_\Sigma = \frac{1}{2} \cdot \frac{d\left(\dfrac{1}{\omega\, L^*(\omega)}\right)}{d\,\omega} U^2 . \tag{4.16}$$

Having written Eqs. (4.14), (4.15) or (4.16) in greater detail, we arrive at the same result:

$$W_\Sigma = \frac{1}{2} C U^2 + \frac{1}{2} L I^2, \tag{4.17}$$

where $U$ is the voltage at the capacitor and $I$ is the current through the inductance coil. Below we consider the physical meaning jog the magnitude $\varepsilon(\omega)$ for material media.

A superconductor is a perfect plasma medium in which charge carriers (electrons) can move without friction. In this case the equation of motion is

$$m \frac{d\vec{V}}{dt} = e\vec{E} , \tag{4.18}$$

where $m$ and $e$ are the electron mass and charge, respectively; $\vec{E}$ is the electric field strength, $\vec{V}$ is the velocity. Taking into account the current density

$$\vec{j} = n e \vec{V}, \tag{4.19}$$

we can obtain from Eq. (4.18)

$$\vec{j}_L = \frac{n e^2}{m} \int \vec{E}\, dt . \tag{4.20}$$

In Eqs. (4.19) and (4.20) $n$ is the specific charge density. Introducing the notion

$$L_k = \frac{m}{n e^2} , \tag{2.21}$$

we can write

$$\vec{j}_L = \frac{1}{L_k}\int \vec{E}\,dt. \qquad (4.22)$$

Here $L_k$ is the kinetic inductivity of the medium. Its existence is based on the fact that a charge carrier has a mass and hence it possesses inertia properties.

For harmonic fields we have $\vec{E}=\vec{E}_0 \sin \omega t$ and Eq. (4.22) becomes

$$\vec{j}_L = -\frac{1}{\omega L_k} E_0 \cos \omega t. \qquad (4.23)$$

Eqs. (4.22) and (4.23) show that $\vec{j}_L$ is the current through the inductance coil.

In this case the Maxwell equations take the following form

$$rot\,\vec{E} = -\mu_0 \frac{\partial \vec{H}}{\partial t},$$
$$rot\,\vec{H} = \vec{j}_C + \vec{j}_L = \varepsilon_0 \frac{\partial \vec{E}}{\partial t} + \frac{1}{L_k}\int \vec{E}\,dt, \qquad (4.24)$$

where $\varepsilon_0$ and $\mu_0$ are the electric and magnetic inductivities in vacuum, $\vec{j}_C$ and $\vec{j}_L$ are the displacement and conduction currents, respectively. As was shown above, $\vec{j}_L$ is the inductive current.

Eq. (4.24) gives

$$rot\,rot\,\vec{H} + \mu_0 \varepsilon_0 \frac{\partial^2 \vec{H}}{\partial t^2} + \frac{\mu_0}{L_k}\vec{H} = 0. \qquad (4.25)$$

For time-independent fields, Eq. (4.25) transforms into the London equation

$$rot\,rot\,\vec{H} + \frac{\mu_0}{L_k}\vec{H} = 0, \qquad (4.26)$$

where $\lambda_L^2 = \frac{L_k}{\mu_0}$ is the London depth of penetration.

As Eq. (4.24) shows, the inductivities of plasma (both electric and magnetic) are frequency – independent and equal to the corresponding parameters for vacuum. Besides, such plasma has another fundamental material characteristic – kinetic inductivity.

Eqs. (4.24) hold for both constant and variable fields. For harmonic fields $\vec{E}=\vec{E}_0 \sin \omega t$, Eq. (4.24) gives

$$rot\,\vec{H} = \left(\varepsilon_0 \omega - \frac{1}{L_k \omega}\right)\vec{E}_0 \cos \omega t. \qquad (4.27)$$

Taking the bracketed value as the specific susceptance $\sigma_x$ of plasma, we can write

$$rot\,\vec{H} = \sigma_X \vec{E}_0 \cos \omega t, \qquad (4.28)$$

where

$$\sigma_X = \varepsilon_0 \omega - \frac{1}{\omega L_k} = \varepsilon_0 \omega \left(1 - \frac{\omega_p^2}{\omega^2}\right) = \omega \, \varepsilon^*(\omega) \, , \tag{4.29}$$

and $\varepsilon^*(\omega) = \varepsilon_0 \left(1 - \frac{\omega_p^2}{\omega^2}\right)$, where $\omega_p^2 = \frac{1}{\varepsilon_0 L_k}$ is the plasma frequency.

Now Eq. (4.28) can be re-written as

$$rot \, \vec{H} = \omega \, \varepsilon_0 \left(1 - \frac{\omega_p^2}{\omega^2}\right) \vec{E}_0 \cos \omega t \, , \tag{4.30}$$

or

$$rot \, \vec{H} = \omega \, \varepsilon^*(\omega) \vec{E}_0 \cos \omega t \, . \tag{4.31}$$

The $\varepsilon^*(\omega)$ –parameter is conventionally called the frequency-dependent electric inductivity of plasma. In reality however this magnitude includes simultaneously the electric inductivity of vacuum aid the kinetic inductivity of plasma. It can be found as

$$\varepsilon^*(\omega) = \frac{\sigma_X}{\omega} \, . \tag{4.32}$$

It is evident that there is another way of writing $\sigma_X$

$$\sigma_X = \varepsilon_0 \omega - \frac{1}{\omega L_k} = \frac{1}{\omega L_k}\left(\frac{\omega^2}{\omega_p^2} - 1\right) = \frac{1}{\omega L_k^*}, \tag{4.33}$$

where

$$L_k^*(\omega) = \frac{L_k}{\left(\frac{\omega^2}{\omega_p^2} - 1\right)} = \frac{1}{\sigma_X \omega} \, . \tag{4.34}$$

$L_k^*(\omega)$ written this way includes both $\varepsilon_0$ and $L_k$.

Eqs. (4.29) and (4.33) are equivalent, and it is safe to say that plasma is characterized by the frequency-dependent kinetic inductance $L_k^*(\omega)$ rather than by the frequency-dependent electric inductivity $\varepsilon^*(\omega)$.

Eq. (4.27) can be re-written using the parameters $\varepsilon^*(\omega)$ and $L_k^*(\omega)$

$$rot \, \vec{H} = \omega \, \varepsilon^*(\omega) \vec{E}_0 \cos \omega t \, , \tag{4.35}$$

or

$$rot \, \vec{H} = \frac{1}{\omega L_k^*(\omega)} \vec{E}_0 \cos \omega t \, . \tag{4.36}$$

Eqs. (4.35) and (4.36) are equivalent.

Thus, the parameter $\varepsilon^*(\omega)$ is not an electric inductivity though it has its dimensions. The same can be said about $L_k^*(\omega)$.

We can see readily that

$$\varepsilon^*(\omega) = \frac{\sigma_X}{\omega}, \qquad (4.37)$$

$$L_k^*(\omega) = \frac{1}{\sigma_X \omega}. \qquad (4.38)$$

These relations describe the physical meaning of $\varepsilon^*(\omega)$ and $L_k^*(\omega)$.

Of course, the parameters $\varepsilon^*(\omega)$ and $L_k^*(\omega)$ are hardly usable for calculating energy by the following equations

$$W_E = \frac{1}{2} \varepsilon\, E_0^2 \qquad (4.39)$$

and

$$W_j = \frac{1}{2} L_k\, j_0^2. \qquad (4.40)$$

For this purpose the Eq. (2.15)-type formula was devised in [9]:

$$W = \frac{1}{2} \cdot \frac{d[\omega\, \varepsilon^*(\omega)]}{d\omega} E_0^2. \qquad (4.41)$$

Using Eq. (4.41), we can obtain

$$W_\Sigma = \frac{1}{2}\varepsilon_0 E_0^2 + \frac{1}{2} \cdot \frac{1}{\omega^2 L_k} E_0^2 = \frac{1}{2}\varepsilon_0 E_0^2 + \frac{1}{2} L_k\, j_0^2. \qquad (4.42)$$

The same result is obtainable from

$$W = \frac{1}{2} \cdot \frac{d\left[\dfrac{1}{\omega L_k^*(\omega)}\right]}{d\omega} E_0^2. \qquad (4.43)$$

As in the case of a parallel circuit, either of the parameters $\varepsilon^*(\omega)$ and $L_k^*(\omega)$, similarly to $C^*(\omega)$ and $L^*(\omega)$, characterize completely the electrodynamic properties of plasma. The case

$$\begin{aligned}\varepsilon^*(\omega) &= 0 \\ L_k^*(\omega) &= \infty\end{aligned} \qquad (4.44)$$

corresponds to the resonance of current.

It is shown below that under certain conditions this resonance can be transverse with respect to the direction of electromagnetic waves.

It is known that the Langmuir resonance is longitudinal. No other resonances have ever been detected in nonmagnetized plasma. Nevertheless, transverse resonance is also possible in such plasma, and its frequency coincides with that of the Langmuir resonance. To understand the origin of the transverse resonance, let us consider a long line consisting of two perfectly conducting planes (see Fig. 4.1). First, we examine this line in vacuum.

If a d.c. voltage ($U$) source is connected to an open line the energy stored in its electric field is

$$W_{E\Sigma} = \frac{1}{2}\varepsilon_0 E^2 a\, b\, z = \frac{1}{2} C_{E\Sigma} U^2, \qquad (4.45)$$

where $E = \dfrac{U}{a}$ is the electric field strength in the line, and

$$C_{E\Sigma} = \varepsilon_0 \frac{b\, z}{a} \qquad (4.46)$$

is the total line capacitance. $C_E = \varepsilon_0 \dfrac{b}{a}$ is the linear capacitance and $\varepsilon_0$ is electric inductivities of the medium (plasma) in SI units (F/m).

The specific potential energy of the electric field is

$$W_E = \frac{1}{2}\varepsilon_0 E^2. \qquad (4.47)$$

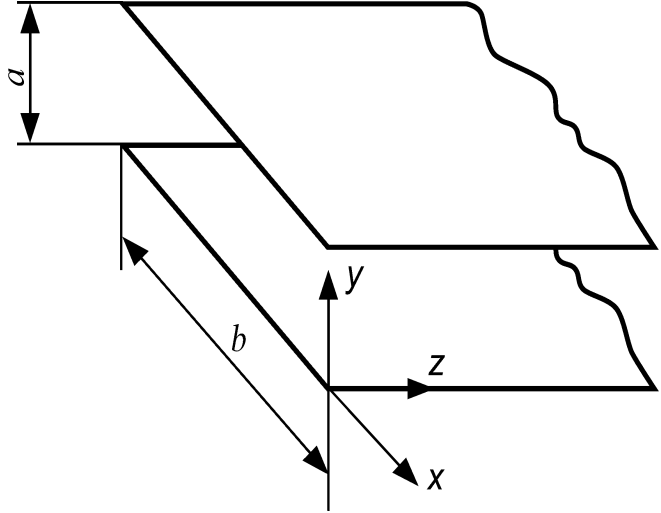

Fig. 4.1. Two-conductor line consisting of two perfectly conducting planes.

If the line is short-circuited at the distance $z$ from its start and connected to a d.c. current ($I$) source, the energy stored in the magnetic field of the line is

$$W_{H\Sigma} = \frac{1}{2}\mu_0 H^2 a\, b\, z = \frac{1}{2} L_{H\Sigma} I^2. \qquad (4.48)$$

Since $H = \dfrac{I}{b}$, we can write

$$L_{H\Sigma} = \mu_0 \frac{a\, z}{b}, \qquad (4.49)$$

where $L_{H\Sigma}$ is the total inductance of the line $L_H = \mu_0 \dfrac{a}{b}$ is linear inductance and $\mu_0$ is the inductivity of the medium (vacuum) in SI (H/m).

The specific energy of the magnetic field is

$$W_H = \frac{1}{2}\mu_0 H^2 . \tag{4.50}$$

To make the results obtained more illustrative, henceforward, the method of equivalent circuits will be used along with mathematical description. It is seen that $C_{E\Sigma}$ and $L_{H\Sigma}$ increase with growing $z$. The line segment $dz$ can therefore be regarded as an equivalent circuit (Fig. 4.2a).

If plasma in which charge carriers can move free of friction is placed within the open line and then the current $I$, is passed through it, the charge carriers moving at a certain velocity start storing kinetic energy. Since the current density is

$$j = \frac{I}{bz} = neV, \tag{4.51}$$

the total kinetic energy of all moving charges is

$$W_{k\Sigma} = \frac{1}{2} \cdot \frac{m}{ne^2} a b z j^2 = \frac{1}{2} \cdot \frac{m}{ne^2} \frac{a}{bz} I^2 . \tag{4.52}$$

On the other hand,

$$W_{k\Sigma} = \frac{1}{2} L_{k\Sigma} I^2 , \tag{4.53}$$

where $L_{k\Sigma}$ is the total kinetic inductance of the line. Hence,

$$L_{k\Sigma} = \frac{m}{ne^2} \cdot \frac{a}{bz} . \tag{4.54}$$

Thus, the magnitude

$$L_k = \frac{m}{ne^2} \tag{4.55}$$

corresponding kinetic inductivity of the medium.

Earlier, we introduced this magnitude by another way (see Eq. (4.21)). Eq. (4.55) corresponds to case of uniformly distributed d.c. current.

As we can see from Eq. (4.54), $L_{H\Sigma}$, unlike $C_{E\Sigma}$ and $L_{k\Sigma}$, decreases when $z$ grows. This is clear physically because the number of parallel-connected inductive elements increases with growing $z$. The equivalent circuit of the line with nondissipative plasma is shown in Fig. 4.2б. The line itself is equivalent to a parallel lumped circuit:

$$C = \frac{\varepsilon_0 b z}{a} \quad \text{and} \quad L = \frac{L_k a}{bz} . \tag{4.56}$$

It is however obvious from calculation that the resonance frequency is absolutely independent of whatever dimension. Indeed,

$$\omega_p^2 = \frac{1}{CL} = \frac{1}{\varepsilon_0 L_k} = \frac{n e^2}{\varepsilon_0 m} \quad . \tag{4.57}$$

This brings us to a very interesting result: the resonance frequency of the macroscopic resonator is independent of its size. It may seem that we are dealing here with the Langmuir resonance because the obtained frequency corresponds exactly to that of the

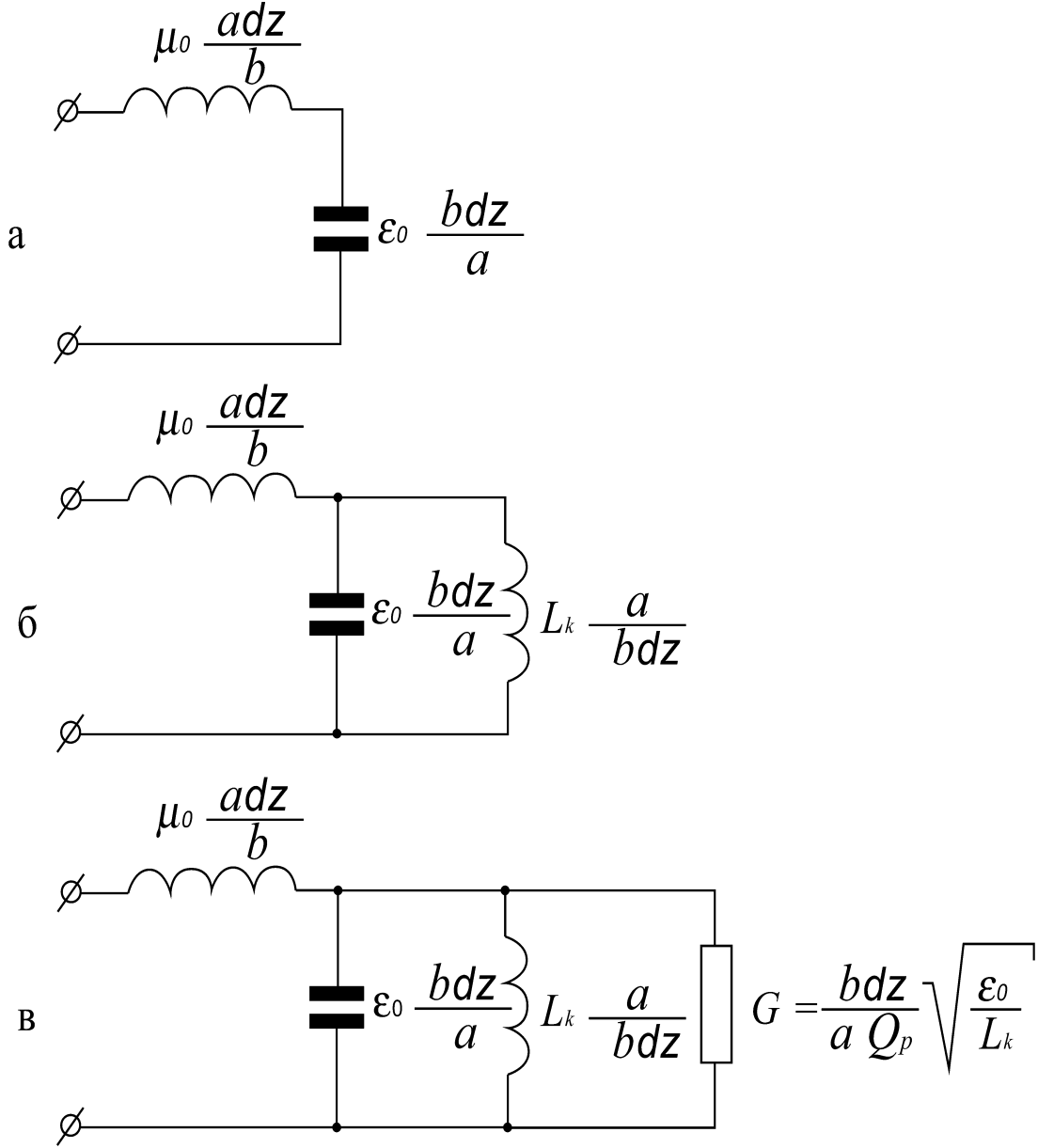

Fig. 4.2. а. Equivalent circuit of the two-conductor line segment;

      б. Equivalent circuit of the two-conductor line segment containing nondissipative plasma;

      в. Equivalent circuit of the two-conductor line segment containing dissipative plasma.

Langmuir resonance. We however know that the Langmuir resonance characterizes longitudinal waves. The wave propagating in the phase velocity in the z-direction is equal to infinity and the wave vector is $\vec{k}_z = 0$, which corresponds to the solution of Eqs. (4.24) for a line of pre-assigned configuration (Fig. 4.1). Eqs. (4.25) give a well-known result. The wave number is

$$k_z^2 = \frac{\omega^2}{c^2}\left(1 - \frac{\omega_p^2}{\omega^2}\right). \tag{4.58}$$

The group and phase velocities are

$$V_g^2 = c^2\left(1 - \frac{\omega_p^2}{\omega^2}\right), \tag{4.59}$$

$$V_F^2 = \frac{c^2}{\left(1 - \frac{\omega_p^2}{\omega^2}\right)}, \tag{4.60}$$

where $c = \left(\frac{1}{\mu_0 \varepsilon_0}\right)^{1/2}$ is the velocity of light in vacuum.

For the plasma under consideration, the phase velocity of the electromagnetic wave is equal to infinity. Hence, the distribution of the fields and currents over the line is uniform at each instant of time and independent of the z-coordinate. This implies that, on the one hand, the inductance $L_{H\Sigma}$ has no effect on the electrodynamic processes in the line and, on the other hand, any two planes can be used instead of conducting planes to confine plasma above and below.

Eqs. (4.58), (4.59) and (4.60) indicate that we have transverse resonance with an infinite Q-factor. The fact of transverse resonance, i.e. different from the Langmuir resonance, is most obvious when the Q-factor is not equal to infinity. Then $k_z \neq 0$ and the transverse wave is propagating in the line along the direction perpendicular to the movement of charge carriers. True, we started our analysis with plasma confined within two planes of a long line, but we have thus found that the presence of such resonance is entirely independent of the line size, i.e. this resonance can exist in an infinite medium. Moreover, in infinite plasma transverse resonance can coexist with the Langmuir resonance characterizing longitudinal waves. Since the frequencies of these resonances coincide, both of them are degenerate. Earlier, the possibility of transverse resonance was not considered. To approach the problem more comprehensively, let us analyze the energy processes in loss-free plasma.

The characteristic resistance of plasma determining the relation between the transverse components of electric and magnetic fields can be found from

$$Z = \frac{E_y}{H_x} = \frac{\mu_0 \omega}{k_z} = Z_0 \left(1 - \frac{\omega_p^2}{\omega^2}\right)^{-1/2}, \qquad (4.61)$$

where $Z_0 = \sqrt{\frac{\mu_0}{\varepsilon_0}}$ is the characteristic resistance in vacuum.

The obtained value of $Z$ is typical for transverse electromagnetic waves in waveguides. When $\omega \to \omega_p$, $Z \to \infty$, and $H_x \to 0$. At $\omega > \omega_p$, both the electric and magnetic field components are present in plasma. The specific energy of the fields is

$$W_{E,H} = \frac{1}{2}\varepsilon_0 E_{0y}^2 + \frac{1}{2}\mu_0 H_{0x}^2. \qquad (4.62)$$

Thus, the energy accumulated in the magnetic field is $\left(1 - \frac{\omega_p^2}{\omega^2}\right)$ times lower than that in the electric field. This traditional electrodynamic analysis is however not complete because it disregards one more energy component – the kinetic energy of charge carriers. It turns out that in addition to the electric and magnetic waves carrying electric and magnetic energy, there is one more wave in plasma – the kinetic wave carrying the kinetic energy of charge carriers. The specific energy of this wave is

$$W_k = \frac{1}{2} L_k j_0^2 = \frac{1}{2} \cdot \frac{1}{\omega^2 L_k} E_0^2 = \frac{1}{2}\varepsilon_0 \frac{\omega_p^2}{\omega^2} E_0^2. \qquad (4.63)$$

The total specific energy thus amounts to

$$W_{E,H,j} = \frac{1}{2}\varepsilon_0 E_{0y}^2 + \frac{1}{2}\mu_0 H_{0x}^2 + \frac{1}{2} L_k j_0^2. \qquad (4.64)$$

Hence, to find the total specific energy accumulated in unit volume of plasma, it is not sufficient to allow only for the fields $E$ and $H$.

At the point $\omega = \omega_p$

$$W_H = 0 \qquad (4.65)$$
$$W_E = W_k,$$

i.e. there is no magnetic field in the plasma, and the plasma is a macroscopic electromechanical cavity resonator of frequency $\omega_p$..

At $\omega > \omega_p$ the wave propagating in plasma carries three types of energy – magnetic, electric and kinetic. Such wave can therefore be-called magnetoelectrokinetic. The kinetic wave is a current-density wave $\vec{j} = \frac{1}{L_k}\int \vec{E}\, dt$. It is shifted by $\pi/2$ with respect to the electric wave.

Up to now we have considered a physically unfeasible case with no losses in plasma, which corresponds to infinite $Q$-factor of the plasma resonator. If losses occur, no matter what physical processes caused them, the $Q$-factor of the plasma resonator is a final quantity. For this case the Maxwell equations become

$$rot\, \vec{E} = -\mu_0 \frac{\partial \vec{H}}{\partial t},$$

$$rot\, \vec{H} = \sigma_{p.ef}\vec{E} + \varepsilon_0 \frac{\partial \vec{E}}{\partial t} + \frac{1}{L_k}\int \vec{E}\, d\, t. \qquad (4.66)$$

The term $\sigma_{p.ef}\vec{E}$ allows for the loss, and the index *ef* near the active conductivity emphasizes that we are interested in the fact of loss and do not care of its mechanism. Nevertheless, even though we do not try to analyze the physical mechanism of loss, we should be able at least to measure $\sigma_{p.ef}$.

For this purpose, we choose a line segment of the length $z_0$ which is much shorter than the wavelength in dissipative plasma. This segment is equivalent to a circuit with the following lumped parameters

$$C = \varepsilon_0 \frac{b\, z_0}{a}, \qquad (4.67)$$

$$L = L_k \frac{d}{b\, z_0}, \qquad (4.68)$$

$$G = \sigma_{p.ef} \frac{b\, z_0}{a}, \qquad (4.69)$$

where $G$ is the conductance parallel to $C$ and $L$.
The conductance $G$ and the $Q$-factor of this circuit are related as

$$G = \frac{1}{Q_\rho}\sqrt{\frac{C}{L}}. \qquad (4.70)$$

Taking into account Eqs. (4.67) – (4.69), we obtain from Eq. (4.70)

$$\sigma_{p.ef} = \frac{1}{Q_\rho}\sqrt{\frac{\varepsilon_0}{L_k}}. \qquad (4.71)$$

Thus, $\sigma_{p.ef}$ can be found by measuring the basic $Q$-factor of the plasma resonator.
Using Eqs. (4.71) and (4.66), we obtain

$$rot\, \vec{E} = -\mu_0 \frac{\partial \vec{H}}{\partial t},$$

$$rot\, \vec{H} = \frac{1}{Q_\rho}\sqrt{\frac{\varepsilon_0}{L_k}}\vec{E} + \varepsilon_0 \frac{\partial \vec{E}}{\partial t} + \frac{1}{L_k}\int \vec{E}\, d\, t. \qquad (4.72)$$

The equivalent circuit of this line containing dissipative plasma is shown in Fig. 4.2b.

Lot us consider the solution of Eqs. (4.72) at the point $\omega = \omega_p$. Since

$$\varepsilon_0 \frac{\partial \vec{E}}{\partial t} + \frac{1}{L_k}\int \vec{E}\, d\, t = 0. \qquad (4.73)$$

We obtain

$$\operatorname{rot} \vec{E} = -\mu_0 \frac{\partial \vec{H}}{\partial t},$$

$$\operatorname{rot} \vec{H} = \frac{1}{Q_P} \sqrt{\frac{\varepsilon_0}{L_k}} \vec{E}.$$
(4.74)

The solution of these equations is well known. If there is interface between vacuum and the medium described by Eqs. (4.74), the surface impedance of the medium is

$$Z = \frac{E_{tg}}{H_{tg}} = \sqrt{\frac{\omega_p \mu_0}{2 \sigma_{p.ef.}}} (1+i),$$
(4.75)

where $\sigma_{p.ef} = \frac{1}{Q_p} \sqrt{\frac{\varepsilon_0}{L_k}}$.

There is of course some uncertainty in this approach because the surface impedance is dependent on the type of the field-current relation (local or non-local). Although the approach is simplified, the qualitative results are quite adequate. True, a more rigorous solution is possible.

The wave propagating deep inside the medium decreases by the law $e^{-\frac{z}{\delta_{ef}}} \cdot e^{-i \frac{z}{\delta_{ef}}}$. In this case the phase velocity is

$$V_F = \omega \, \sigma_{p.ef},$$
(4.76)

where $\delta_{p.ef}^2 = \frac{2}{\mu_0 \omega_p \sigma_{p.ef}}$ is the effective depth of field penetration in the plasma. The above relations characterize the wave process in plasma. For good conductors we usually have $\frac{\sigma_{ef}}{\omega \varepsilon_0} \gg 1$. In such a medium the wavelength is

$$\lambda_g = 2\pi\delta.$$
(4.77)

I.e. much shorter than the free-space wavelength. Further on we concentrate on the case $\lambda_g \gg \lambda_0$ at the point $\omega = \omega_p$, i.e. $V_F \big|_{\omega = \omega p} \gg c$.

We have found that $\varepsilon(\omega)$ is not dielectric inductivity permittivity. Instead, it includes two frequency-independent parameters $\varepsilon_0$ and $L_k$. What is the reason for the physical misunderstanding of the parameter $\varepsilon(\omega)$? This occurs first of all because for the case of plasma the $\frac{1}{L_k} \int \vec{E} \, dt$ - type term is not explicitly present in the second Maxwell equation.

There is however another reason for this serious mistake in the present-day physics [9] as an example. This study states that there is no difference between di-electrics and conductors at very high frequencies. On this basis the authors suggest the existence of a polarization vector in conducting media and this vector is introduced from the relation

$$\vec{P} = \Sigma\, e\, \vec{r}_m = n\, e\, \vec{r}_m, \qquad (4.78)$$

where $n$ is the charge carrier density, $\vec{r}_m$ is the current charge displacement. This approach is physically erroneous because only bound charges can polarize and form electric dipoles when the external field overcoming the attraction force of the bound charges accumulates extra electrostatic energy in the dipoles. In conductors the charges are not bound and their displacement would not produce any extra electrostatic energy. This is especially obvious if we employ the induction technique to induce current (i.e. to displace charges) in a ring conductor. In this case there is no restoring force to act upon the charges, hence, no electric polarization is possible. In [9] the polarization vector found from Eq. (4.78) is introduced into the electric induction of conducting media

$$\vec{D} = \varepsilon_0\, \vec{E} + \vec{P}, \qquad (4.79)$$

where the vector $\vec{P}$ of a metal is obtained from Eq. (4.78), which is wrong.

Since

$$\vec{r}_m = -\frac{e^2}{m\,\omega^2}\vec{E}, \qquad (4.80)$$

for free carriers, then

$$\vec{P}^*(\omega) = -\frac{n\,e^2}{m\,\omega^2}\vec{E}, \qquad (4.81)$$

for plasma, and

$$\vec{D}^*(\omega) = \varepsilon_0\,\vec{E} + \vec{P}^*(\omega) = \varepsilon_0\left(1 - \frac{\omega_p^2}{\omega^2}\right)\vec{E}. \qquad (4.82)$$

Thus, the total accumulated energy is

$$W_\Sigma = \frac{1}{2}\varepsilon_0\, E^2 + \frac{1}{2}\cdot\frac{1}{L_k\,\omega^2}E^2. \qquad (4.83)$$

However, the second term in the right-hand side of Eq. (4.83) is the kinetic energy (in contrast to dielectrics for which this term is the potential energy). Hence, the electric induction vector $D^*(\omega)$ does not correspond to the physical definition of the electric induction vector.

The physical meaning of the introduced vector $\vec{P}^*(\omega)$ is clear from

$$\vec{P}^*(\omega) = \frac{\sigma_L}{\omega}\vec{E} = \frac{1}{L_k\,\omega^2}\vec{E}. \qquad (4.84)$$

The interpretation of $\varepsilon(\omega)$ as frequency-dependent inductivity has been harmful for correct understanding of the real physical picture (especially in the educational processes). Besides, it has drawn away the researchers attention from some physical phenomena in plasma, which first of all include the transverse plasma resonance and three energy components of the magnetoelectrokinetic wave propagating in plasma.

Below, the practical aspects of the results obtained are analyzed, which promise new data and refinement of the current views.

Plasma can be used first of all to construct a macroscopic single-frequency cavity for development of a new class of electrokinetic plasma lasers. Such cavity can also operate as a band-pass filter.

At high enough $Q_p$ the magnetic field energy near the transverse resonance is considerably lower than the kinetic energy of the current carriers and the electrostatic field energy. Besides, under certain conditions the phase velocity can much exceed the velocity of light. Therefore, if we want to excite the transverse plasma resonance, we can put

$$rot \vec{E} \cong 0,$$

$$\frac{1}{Q_p}\sqrt{\frac{\varepsilon_0}{L_k}}\vec{E} + \varepsilon_0 \frac{\partial \vec{E}}{\partial t} + \frac{1}{L_k}\int \vec{E}\, dt = \vec{j}_{CT}, \qquad (4.85)$$

where $\vec{j}_{CT}$ is the extrinsic current density.

Integrating Eq. (2.84) over time and dividing it by $\varepsilon_0$ obtain

$$\omega_p^2 \vec{E} + \frac{\omega_p}{Q_p}\cdot\frac{\partial \vec{E}}{\partial t} + \frac{\partial^2 \vec{E}}{\partial t^2} = \frac{1}{\varepsilon_0}\cdot\frac{\partial \vec{j}_{CT}}{\partial t}. \qquad (4.86)$$

Integrating Eq. (4.86) over the surface normal to the vector $\vec{E}$ and taking $\Phi_E = \int \vec{E}\, d\vec{S}$, we have

$$\omega_p^2 \Phi_E + \frac{\omega_p}{Q_p}\cdot\frac{\partial \Phi_E}{\partial t} + \frac{\partial^2 \Phi_E}{\partial t^2} = \frac{1}{\varepsilon_0}\cdot\frac{\partial I_{CT}}{\partial t}, \qquad (4.87)$$

where $I_{CT}$ is the extrinsic current.

Eq. (4.87) is the harmonic oscillator equation whose right-hand side is typical of two-level lasers [10]. If there is no excitation source, we have a "cold". Laser cavity in which the oscillation damping follows the exponential law

$$\Phi_E(t) = \Phi_E(0) e^{i\omega_p t} \cdot e^{-\frac{\omega_p}{2Q_p}t}, \qquad (4.88)$$

i.e. the macroscopic electric flow $\Phi_E(t)$ oscillates at the frequency $\omega_p$. The relaxation time can be round as

$$\tau = \frac{2Q_P}{\omega_P}. \qquad (4.89)$$

If this cavity is excited by extrinsic currents, the cavity will operate as a band-pass filter with the pass band $\Delta\omega = \dfrac{\omega_p}{2Q_p}$.

Transverse plasma resonance offers another important application – it can be used to heat plasma. High-level electric fields and, hence, high change-carrier energies can be obtained in the plasma resonator if its $Q$-factor is high, which is achievable at low concentrations of plasma. Such cavity has the advantage that the charges attain the highest velocities far from cold planes. Using such charges for nuclear fusion, we can keep the process far from the cold elements of the resonator.

Such plasma resonator can be matched easily to the communication line. Indeed, the equivalent resistance of the resonator at the point $\omega = \omega_p$ is

$$R_{\text{экв}} = \frac{1}{G} = \frac{a\, Q_P}{b\, z} \sqrt{\frac{L_k}{\varepsilon_0}}. \tag{4.90}$$

The communication lines of sizes $a_L$ and $b_L$ should be connected to the cavity either through a smooth junction or in a stepwise manner. If $b = b_L$, the matching requirement is

$$\frac{a_L}{b_L} \sqrt{\frac{\mu_0}{\varepsilon_0}} = \frac{a\, Q_p}{b\, z_0} \sqrt{\frac{L_k}{\varepsilon_0}}, \tag{4.91}$$

$$\frac{a\, Q_p}{a_L z_0} \sqrt{\frac{L_k}{\mu_0}} = 1. \tag{4.92}$$

It should be remembered that the choice of the resonator length $z_0$ must comply with the requirement $z_0 \ll \lambda_g \big|_{\omega = \omega_p}$.

Development of devices based on plasma resonator can require coordination of the resonator and free space. In this case the following condition is important:

$$\sqrt{\frac{\mu_0}{\varepsilon_0}} = \frac{a\, Q_p}{b\, z_0} \sqrt{\frac{L_k}{\varepsilon_0}}, \tag{4.93}$$

or

$$\frac{a\, Q_p}{b\, z_0} \sqrt{\frac{L_k}{\mu_0}} = 1. \tag{4.94}$$

Such plasma resonators can be excited with d.c. current, as is the case with a monotron microwave oscillator [11]. It is known that a microwave diode (the plasma resonator in our case) with the transit angle of $\sim 5/2\pi$ develops negative resistance and tends to self-excitation. The requirement of the transit angle equal to $5/2\pi$ correlates with the following d.c. voltage applied to the resonator:

$$U_0 = \frac{0{,}32 a^2\, \omega_p^2\, m\, c^2}{4\pi^2\, e} = \frac{0{,}32 a^2\, n\, e}{4\pi^2\, \varepsilon_0^2\, \mu_0}, \tag{4.95}$$

where $a$ is the distance between the plates in the line.

It is quite probable that this effect is responsible for the electromagnetic oscillations in semiconductive lasers.

## 4.2. Dielectric media.

Applied fields cause polarization of bound charges in dielectrics. The polarization takes some energy from the field source, and the dielectric accumulates extra electrostatic energy. The extent of displacement of the polarized charges from the equilibrium is dependent on the electric field and the coefficient of elasticity β, characterizing the elasticity of the charge bonds. These parameters are related as

$$-\omega^2 \vec{r}_m + \frac{\beta}{m}\vec{r}_m = \frac{e}{m}\vec{E}, \qquad (4.96)$$

where $\vec{r}_m$ is the charge displacement from the equilibrium.

Putting $\omega_0$ for the resonance frequency of the bound charges and taking into account that $\omega_0 = \beta/m$ we obtain from Eq. (4.96)

$$\vec{r}_m = -\frac{e\vec{E}}{m(\omega^2 - \omega_o^2)} \qquad (4.97)$$

The polarization vector becomes

$$\vec{P}_m^* = -\frac{n e^2}{m} \cdot \frac{1}{(\omega^2 - \omega_0^2)} \vec{E}. \qquad (4.98)$$

Since

$$\vec{P} = \varepsilon_0 (\varepsilon - 1) \vec{E}, \qquad (4.99)$$

we obtain

$$\varepsilon'_\partial{}^*(\omega) = 1 - \frac{n e^2}{\varepsilon_0 m} \cdot \frac{1}{\omega^2 - \omega_0^2}. \qquad (4.100)$$

The quantity $\varepsilon'_\partial{}^*(\omega)$ is commonly called the relative frequency dependably lectric inductivity. Its absolute value can be found as

$$\varepsilon_\partial{}^*(\omega) = \varepsilon_0 (1 - \frac{n e^2}{\varepsilon_0 m} \cdot \frac{1}{\omega^2 - \omega_0^2}). \qquad (4.101)$$

Once again, we arrive at the frequency-dependent dielectric permitlivity. Let us take a closer look at the quantity $\varepsilon_\partial{}^*(\omega)$. As before, we introduce $L_{k\,\partial} = \frac{m}{n e^2}$ and $\omega_{p.\partial} = \dfrac{1}{L_{k\,\partial}\varepsilon_0}$ and see immediately that the vibrating charges of the dielectric have masses and thus possess inertia properties. As a result, their kinetic inductivity would make itself evident too. Eq. (4.101) can be re-written as

$$\varepsilon_\partial *(\omega) = \varepsilon_0 (1 - \dfrac{\omega_{p\,\partial}^2}{\omega^2 - \omega_0^2}). \qquad (4.102)$$

It is appropriate to examine two limiting cases: $\omega \gg \omega_0$ and $\omega \ll \omega_0$.

If $\omega \gg \omega_0$,

$$\varepsilon_\partial *(\omega) = \varepsilon_0 (1 - \dfrac{\omega_{p\,\partial}^2}{\omega^2}), \qquad (4.103)$$

and the dielectric behaves just like plasma. This case has prompted the idea that at high frequencies there is no difference between dielectrics and plasma. The idea served as a basis for introducing the polarization vector in conductors [9]. The difference however exists and it is of fundamental importance. In dielectrics, because of inertia, the amplitude of charge vibrations is very small at high frequencies and so is the polarization vector. The polarization vector is always zero in conductors.

For $\omega \ll \omega_0$,

$$\varepsilon_\partial *(\omega) = \varepsilon_0 (1 + \dfrac{\omega_{p\,\partial}^2}{\omega_0^2}), \qquad (4.104)$$

and the permittivity of the dielectric is independent of frequency. It is $(1 + \dfrac{\omega_{p\,\partial}^2}{\omega_0^2})$ times higher than in vacuum. This result is quite clear. At $\omega \gg \omega_0$ the inertia properties are inactive and permittivity approaches its value in the static field.

The equivalent circuits corresponding to these two cases are shown in Figs. 4.3a and b. It is seen that in the whole range of frequencies the equivalent circuit of the dielectric acts as a series oscillatory circuit parallel-connected to the capacitor operating due to the electric inductivity $\varepsilon_0$ of vacuum (see Fig. 4.3b). The resonance frequency of this series circuit is obviously obtain from

$$\omega_\partial^2 = \dfrac{1}{L_k \varepsilon_0 \left(\dfrac{\omega_{p\,\partial}^2}{\omega_0^2}\right)}. \qquad (4.105)$$

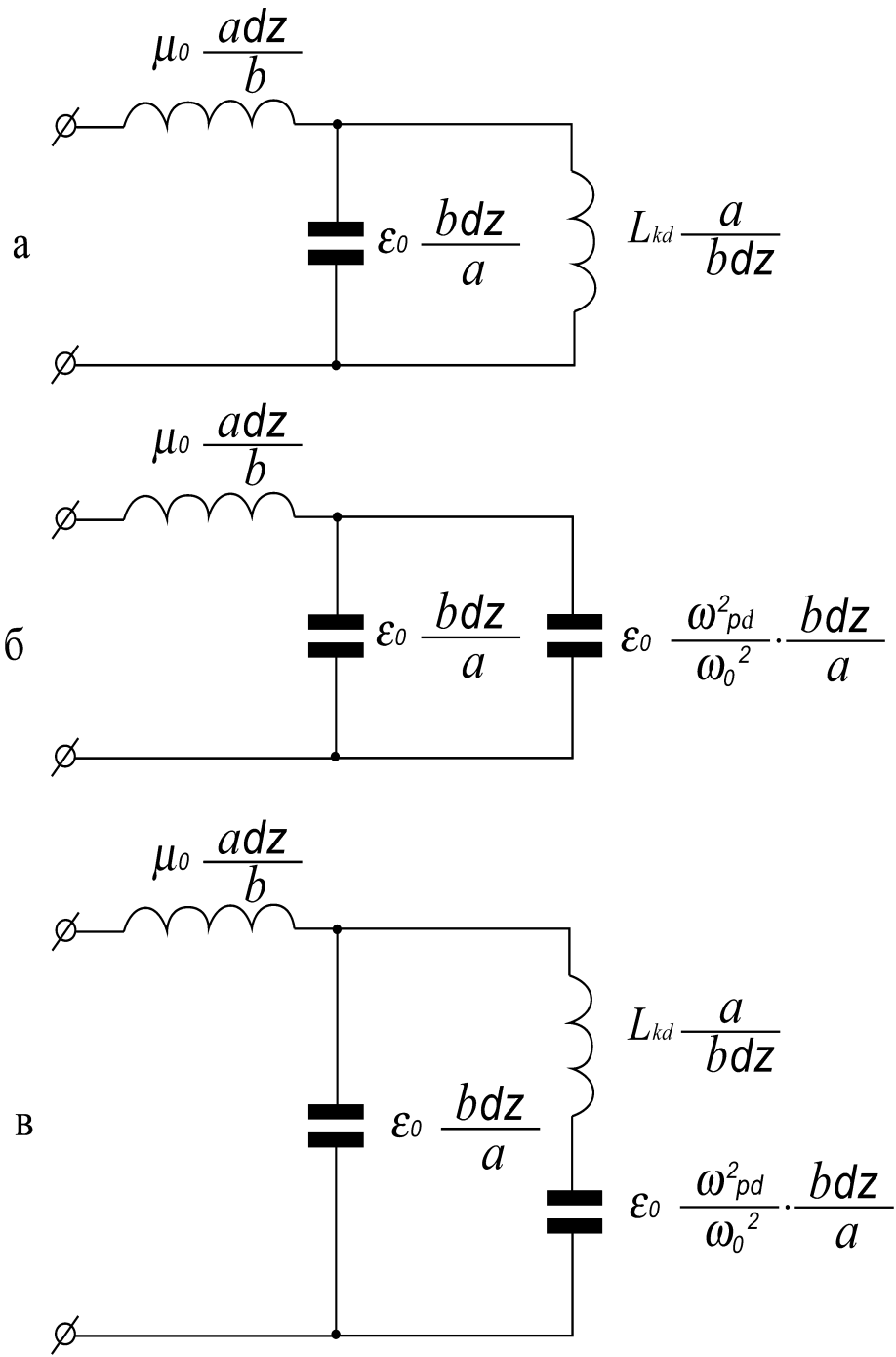

Fig. 4.3. Equivalent circuit of two-conductor line segment with a dielectric: а - $\omega \gg \omega_0$; б – $\omega \ll \omega_0$; в – the whole frequency range.

Lake in the case of plasma, $\omega_0^2$ is independent of the line size, i.e. we have a macroscopic resonator whose frequency is only true when there are no bonds between individual pairs of bound charges.

Like for plasma, $\varepsilon_\partial*(\omega)$ is specific susceptance of the dielectric divided by frequency. However, unlike plasma, this parameter contains three frequency-independent components: $\varepsilon_0$, $L_{k\partial}$ and the static permittivity of the dielectric $\varepsilon_0 \dfrac{\omega_{p\partial}^2}{\omega_0^2}$. In the dielectric, resonance occurs when $\varepsilon_\partial*(\omega) \to -\infty$.

Three waves-magnetic, electric and kinetic-propagate in it too. Each of them carries its own type of energy. It not is not problematic to calculate them but we omit this here to save room.

### 4.3. Magnetic media.

The resonance phenomena in plasma and dielectrics are characterized by repeated electrostatic-kinetic and kinetic-electrostatic transformations of the charge motion energy during oscillations. This can be described as an electrokinetic process, and devices based on it (lasers, masers, filters, etc.) can be classified as electrokinetic units.

However, another type of resonance is also possible, namely, magnetic resonance. Within the current concepts of frequency-dependent permeability, it is easy to show that such dependence is related to magnetic resonance. For example, let us consider ferromagnetic resonance. A ferrite magnetized by applying a stationary field $H_0$ parallel to the z-axis will act as an anisotropic magnet in relation to the variable external field. The complex permeability of this medium has the form of a tensor [12]:

$$\mu = \begin{pmatrix} \mu_T*(\omega) & -i\alpha & 0 \\ i\alpha & \mu_T*(\omega) & 0 \\ 0 & 0 & \mu_L \end{pmatrix}, \quad (4.106)$$

where

$$\mu_T*(\omega) = 1 - \frac{\Omega|\gamma|M_0}{\mu_0(\omega^2-\Omega^2)}, \quad \alpha = \frac{\omega|\gamma|M_0}{\mu_0(\omega^2-\Omega^2)}, \quad \mu_L = 1, \quad (4.107)$$

$$\Omega = |\gamma|H_0. \quad (4.108)$$

Being the natural professional frequency, and

$$M_0 = \mu_0(\mu-1)H_0 \quad (4.109)$$

is the medium magnetization.

Taking into account Eqs. (4.108) and (4.109) for $\mu_T*(\omega)$, we can write

$$\mu_T*(\omega) = 1 - \frac{\Omega^2(\mu-1)}{\omega^2-\Omega^2}. \quad (4.110)$$

Assuming that the electromagnetic wave propagates along the *x*-axis and there are $H_y$ and $H_z$ components, the first Maxwell equation becomes

$$rot\, \vec{E} = \frac{\partial \vec{E}_z}{\partial x} = \mu_0 \mu_T \frac{\partial \vec{H}_y}{\partial t} \quad . \tag{4.111}$$

Taking into account Eq. (4.110), we obtain

$$rot\, \vec{E} = \mu_0 \left[1 - \frac{\Omega^2(\mu-1)}{\omega^2 - \Omega^2}\right] \frac{\partial \vec{H}_y}{\partial t} \quad . \tag{4.112}$$

For $\omega \gg \Omega$

$$rot\, \vec{E} = \mu_0 \left[1 - \frac{\Omega^2(\mu-1)}{\omega^2}\right] \frac{\partial \vec{H}_y}{\partial t} \quad . \tag{4.113}$$

Assumeng $\vec{H}_y = \vec{H}_{y0} \sin \omega t$ and taking into account that

$$\frac{\partial \vec{H}_y}{\partial t} = -\omega^2 \int \vec{H}_y\, dt \quad . \tag{4.114}$$

Eq. (4.113) gives

$$rot\, \vec{E} = \mu_0 \frac{\partial \vec{H}_y}{\partial t} + \mu_0 \Omega^2 (\mu-1) \int \vec{H}_y\, dt \quad , \tag{4.115}$$

or

$$rot\, \vec{E} = \mu_0 \frac{\partial \vec{H}_y}{\partial t} + \frac{1}{C_k} \int \vec{H}_y\, dt \quad . \tag{4.116}$$

For $\omega \ll \Omega$

$$rot\, \vec{E} = \mu_0 \mu \frac{\partial \vec{H}_y}{\partial t} \quad . \tag{4.117}$$

The quantity

$$C_k = \frac{1}{\mu_0 \Omega^2 (\mu-1)} \tag{4.118}$$

can be described as kinetic capacitance. What is its physical meaning? If the direction of the magnetic moment does not coincide with that of the external magnetic field, the vector of the moment starts precessional motion at the frequency $\Omega$ about the magnetic field vector. The magnetic moment $\vec{m}$ has the potential energy $U_m = -\vec{m} \cdot \vec{B}$. Like in a charged condenser, $U_m$ is the potential energy because the precessional motion is inertialess (even though it is mechanical) and it stops immediately when the magnetic field is lifted. In the magnetic field the processional motion lasts until the accumulated potential energy is exhausted and the vector of the magnetic moment becomes parallel to the vector $\vec{H}_0$.

The equivalent circuit for this case is shown in Fig. 4.4. Magnetic resonance occurs at the point ω=Ω and $\mu_r^*(\omega) \to -\infty$. It is seen that the resonance frequency of the macroscopic magnetic resonator is independent of the line size and equals Ω.

Thus, the parameter

$$\mu_H^*(\omega) = \mu_0 \left[ 1 - \frac{\Omega^2(\mu-1)}{\omega^2 - \Omega^2} \right] \qquad (4.119)$$

is not a frequency-dependent permeability. According to the equivalent circuit in Fig. 4.4, it includes $\mu_0$, $\mu$ and $C_k$

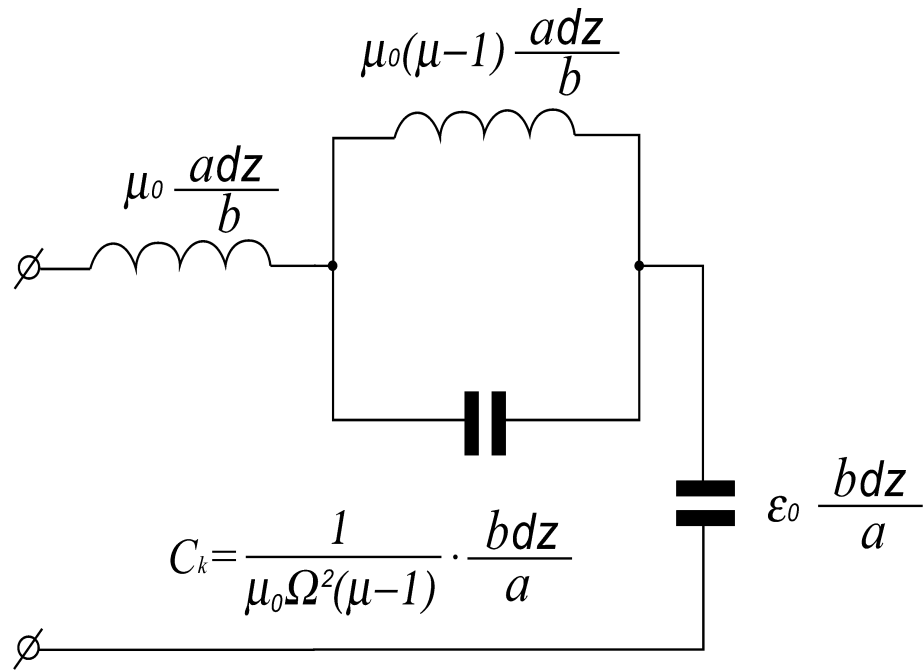

Рис. 4.4. Equivalent circuit of two-conductor line including a magnet.

It is easy to show that three waves propagate in this case-electric, magnetic and a wave carrying potential energy of the precessional motion of the magnetic moments about the vector $\vec{H}_0$. The systems in which these types of waves are used can also be described as electromagnetopotential devices.

## 5. Conclusions

Thus, it has been found that along with the fundamental parameters $\varepsilon\varepsilon_0$ and $\mu\mu_0$ characterizing the electric and magnetic energy accumulated and transferred in the medium, there are two more basic material parameters $L_k$ and $C_k$. They characterize kinetic and potential energy that can be accumulated and transferred in mate-rial media. $L_k$ was sometimes used to describe certain physical phenomena, for example, in superconductors [13], $C_k$ has never been known to exist. These four fundamental parameters $\varepsilon\varepsilon_0$, $\mu\mu_0$, $L_k$ and $C_k$ clarify the physical picture of the wave and resonance processes in material media in applied electromagnetic fields. Previously, only electromagnetic waves were thought to propagate and transfer energy in material media. It is clear now that the concept was not complete. In fact, magnetoelectrokinetic, or electromagnetopotential waves travel in material media. The resonances in these media also have specific features. Unlike closed planes with electromagnetic resonance and energy exchange between electric and magnetic fields, material media have two types of resonance – electrokinetic and magnetopotential. Under the electrokinetic resonsnce the energy of the electric field changes to kinetic energy. In the case of magnetopotential resonance the potential energy accumulated during the precessional motion can escape outside at the precession frequency.

The notions of permittivity and permeability dispersion thus become physically groundless though $\varepsilon*(\omega)$ and $\mu*(\omega)$ are handy for a mathematical description of the processes in material media. We should however remember their true meaning especially where educational processes are involved.

This paper is an improved version of paper [14], the importance of introduction of the scalar-vector potential is demonstrated here more clearly.

The author is indebted to V.D.Fil for helpful discussions, to N.P. Mende and A.I.Shurupov for their assistance in preparation of this manuscript.